\documentclass{aa}  
\usepackage{keyval}
\usepackage{graphicx}
\usepackage{txfonts}
\usepackage{lipsum}
\usepackage{subcaption}         
                               
\usepackage{lscape}             
                                
\usepackage{placeins}

\usepackage{hyperref}
\hypersetup{
    colorlinks=true,
    linkcolor=blue,
    filecolor=magenta,      
    urlcolor=blue,
    citecolor=blue,
}
\usepackage{xcolor}

\titlerunning{Coronal Current Sheets with Guide-Fields}

\begin{document}

   \title{Dynamical and thermal evolution of localized reconnecting current sheets under varying guide-fields in the solar corona}

   \author{Saksham Pande\inst{1,2
   }\fnmsep\thanks{Corresponding author: \href{mailto:saksham.pande@kuleuven.be}{saksham.pande@kuleuven.be}}
        \and Abhishekh Kumar Srivastava\inst{2}
        \and Fabio Bacchini\inst{1,3}
        \and Sripan Mondal\inst{2}
        \and Akash Bairagi\inst{2}
        \and Balveer Singh\inst{4}
        \and Kartika Sangal\inst{5}
        \and Sudheer K. Mishra\inst{6}
        \and Xinping ~Zhou\inst{7}
        \and Ding~Yuan\inst{5}
        \and R.-Y.~Kwon\inst{8}
        \and B.N.~Dwivedi\inst{2}
        }

   \institute{Centre for mathematical Plasma Astrophysics (CmPA), KU Leuven, 3001 Leuven, Belgium
   \and Department of Physics, Indian Institute of Technology (BHU), Varanasi-221005, India
   \and Royal Belgian Institute for Space Aeronomy, Solar-Terrestrial Centre of Excellence, 1180 Uccle, Belgium
   \and Division of Mathematics, School of Science and Engineering, University of Dundee, United Kingdom
   \and Key Laboratory of Solar Activity and Space Weather, School of Aerospace, Harbin Institute of Technology, Shenzhen, Guangdong 518055, China
   \and Astronomical Observatory, Kyoto University, Japan
   \and College of Physics and Electronic Engineering, Sichuan Normal University, Chengdu 610068, People's Republic of China
   \and KASI, Daejeon, S. Korea
      }

   \date{  }

  \abstract
   {Although magnetic reconnection in coronal current sheets is a fundamental energy release mechanism in the solar corona, the role of the guide-field in governing its formation and self-consistent dynamical and thermal evolution is not fully understood.}
   {We investigate how varying the guide-field strength affects the formation and evolution of a localized reconnecting coronal current sheet, as well as its thinning, reconnection rate, and plasma properties.} 
   {We perform a 2.5-dimensional resistive magnetohydrodynamic (MHD) simulations of the formation and subsequent reconnection of coronal current sheets across a diverse range of guide-field strengths, tracking the temporal evolution of the reconnection rate, current-sheet geometry, temperature, and outflow velocities.}
   {A stronger guide-field prevents oppositely directed field lines from approaching sufficiently to form the current sheet, delaying and suppressing its subsequent evolution. The reconnection rate ($\eta J_{max}$) exhibits a rising phase resembling a power-law, with exponent $\delta_1$ decreasing with increasing guide-field, indicating slower temporal growth of the reconnection rate with increasing guide-fields. The aspect ratio, which characterizes the thinning of the current sheet, follows a similar power-law trend with the exponent $\alpha_1$ also decreasing with the strength of the guide-field. A subsequent slower growth phase of $\eta J_{max}$ shows a power-law exponent $\delta_2$ that increases with guide-field, as does the corresponding aspect-ratio exponent $\alpha_{2}$. During the evolution of the current sheet, the latter phase is achieved due to the sharp decremental transition of the guide-field within the current sheet after a time  when the increasing inflows are inefficiently resisted by the guide-field itself. In the reconnection regime during prior phases, the power-law exponents of both the maximum temperature and the outflow velocities also decrease with increasing guide-field demonstrating that stronger guide-fields suppress the growth of these observational signatures of reconnection.  
   }
   {The presented study suggests that in the presence of non-zero guide fields, a self-consistently formed current sheet displays phases of efficient and inefficient quenching of the reconnection mechanism, promoted by the aforementioned guide-fields. With increasing guide-field strengths, we further quantify this behavior for various indicators of the reconnection process. We also discuss the applicability of the obtained results to observations of apparent current sheets as well as probing background coronal conditions without any knowledge of guide-field observations.
   
   }


   \maketitle
   \nolinenumbers

\section{Introduction}
Magnetic reconnection is a fundamental process in resistive magnetized plasmas during which magnetic-field lines of nearly opposite directions exchange their topological connectivity, resulting in the release of stored magnetic energy in the form of thermal energy, kinetic energy, and particle acceleration at diverse spatial scales such as astrophysical and laboratory systems \citep[e.g.,][and references therein]{priest2014magnetohydrodynamics, pontin2022magnetic}. It is a popular candidate for explaining a wide range of dynamic phenomena in the solar corona, such as flares, coronal mass ejections (CMEs), jets, and coronal heating, as reported by several observations \citep[e.g.,][and references therein]{hudson2011global,webb2012coronal,benz2017flare,cheng2020initiation,gopalswamy2024soho}. The physical processes underlying reconnection have also been investigated in many numerical experiments \citep[e.g.,][and references therein]{birn2005forced,shay2014electron,zweibel2016perspectives,pontin2022magnetic}. The most widely accepted indicator of reconnection in the solar atmosphere is the presence of a narrow diffusion region between antiparallel field lines, which is described as the current sheet (CS).
The CS lies within the magnetic topology on a much smaller length scale than the global length scales of observations and simulations. In imaging observations, the CS is observed as a bright ray-like structure or as an apparent current sheet \citep[e.g.,][and references therein]{xue2020solar,ding2022magnetic,yan2022fast,mondal2023reconnection,2025ApJ...984...36S}. In two dimensions, reconnection can only take place at the magnetic X-point located within such a CS.

The most basic reconnection system involves the encounter and diffusion of the two straight flux tubes possessing antiparallel magnetic fields. In principle, the magnetic reconnection takes place under complex situations in the 3-D corona, either owing to the oblique orientation of one flux tube with respect to the other flux tube, or as a result of the twist of flux tubes along their orientation axis. Therefore, deviations occur from simplified past theoretical and numerical models \citep[e.g.,][]{parker1957sweet}, where analysis was limited to the reconnection between a set of antiparallel magnetic field lines (i.e., the components of the incoming magnetic field lines belong to the planar geometry). The magnetic fields in the natural eruptive events \citep[e.g.,][]{trattner2007location,qiu2010reconnection} may be decomposed into its antiparallel planar components and an additional magnetic-field component perpendicular to it called the guide-field. A well established example is that of the reconnection in the solar arcade loops \citep[e.g.,][]{moore2001onset}. There exists literature suggesting the omnipresence of guide fields in the quiet Sun (QS) \citep[e.g,][]{dahlin2022variability,robinson2022incoherent} as well as in the active regions (ARs) \citep[e.g,][]{qiu2010reconnection,chen2020measurement} of the corona. Indirect measurements of guide-fields using flare observations suggest that varying strengths with respect to the in-flowing \citep[][]{qiu2010reconnection} or out-flowing magnetic field \citep[][]{chen2020measurement} can be assigned to the non-reconnecting component. However, the occurrence of magnetic reconnection is reported both in QS and ARs, indicating that such fields may strongly influence the reconnection process.

The role of the guide-field in magnetic reconnection has been examined through a wide range of theoretical, numerical, observational, and laboratory studies. For a strong guide-field, particularly when its strength is comparable to that of the antiparallel in-plane magnetic fields, several simulation studies have demonstrated that the reconnection rate is reduced \citep[e.g.,][]{pritchett2004three,huba2005hall} with respect to the case of zero guide-field. Laboratory measurements have also confirmed that the reconnection rate can be significantly diminished with an increasing guide-field, often accompanied by changes in the Hall field structure and pressure balance within and downstream of the reconnection region \citep[][]{Tharp2012}. Additionally, observational and laboratory studies have highlighted that the guide-field may play a crucial role in determining the internal structure of the reconnecting layer, particle energization, and force balance in the diffusion region \citep[][]{qiu2010reconnection}. For a weaker guide-field, the reconnection rate may not be strongly affected \citep[e.g.,][]{huba2005hall, birn2010energy,huang2016turbulent}.  

Many of these studies consider preexisting Harris-type current sheets. Assuming a preformed CS does not capture its self-consistent formation and evolution, which, in driven systems, occurs progressively over a finite interval after the driver is switched on. Consequently, how the guide-field modulates not only the reconnection rate but also the timing and morphology of CS formation, the properties of bidirectional outflows, and the associated temperature enhancements, remains insufficiently explored. Therefore, it is important to systematically investigate how the guide-field strength affects the formation and subsequent evolution of the CS. A complete three dimensional system would allow the reconnection-generated flows and energy to escape from the reconnection source region, and allow for a wider class of instabilities and turbulent flows due to the generation of magnetic islands along separatrices \citep{schreier2010three, daughton2011role}. The inclusion and analysis of three-dimensional effects such as oblique tearing modes and out-of-plane instabilities \citep{beg2022} are deferred to future work.

Flux cancellation is a well-known driver for reconnection. In this mechanism, magnetic fragments of opposite polarity converge towards each other when subjected to converging photospheric motions, which also results in the movements of the coronal magnetic-field topology anchored to these polarities, further promoting magnetic reconnection between oppositely directed field lines. 
An enhanced rate of cancellation, as suggested by observations \citep[][]{solanki2010sunrise,solanki2017second,smitha2017estimation} puts the spotlight on its potential role as a key contributor to chromospheric and coronal heating, which led \citet{priest2018cancellation} and \citet{syntelis2019cancellation} to propose photospheric-reconnection driven by the approach of two oppositely
directed photospheric magnetic sources and the subsequent cancellation of their converging fluxes in the presence of an overlying field that is taken to be horizontal and constant for simplicity. However, they have not included any magnetic-field component normal to the plane of reconnection, i.e., a guide-field, in their two-dimensional numerical setup. The 2.5D approximation enforces translational symmetry in the $z$-direction while retaining all three vector components of $\mathbf{B}$ and $\mathbf{v}$. This allows the guide-field component $B_z$ to evolve self-consistently and enter the in-plane force balance as a magnetic pressure term, which is absent in purely 2D treatments. It should be noted that many times converging motions of the magnetic polarities and their subsequent response in the overlying corona may not lead to flux-cancellation but rearrange the localized corona leading to the formation of CS, subsequent reconnection, and generation of complex wave dynamics as well as plasma flows \citep{2025ApJ...989..222M}.

In this work, we study the spatial and temporal evolution of the CS formed due to the collapse of X-type null points caused by the motion of  field lines anchored to these opposing magnetic polarities towards each other at the coronal base. This is done in the absence, as well as in the presence, of weak to strong guide fields. The setup is very similar to the flux cancellation scenario, but the rearrangements of the coronal field lines lead to CS formation and reconnection processes instead of the dissipation of the magnetic fluxes directly at the footpoints. We also explore the role of guide-fields on the reconnection rate and on how the evolving physical conditions of the reconnection regions vary over the time. This is done to identify trends in CS properties and reconnection indicators as a function of varying guide-field strength. The 2.5D resistive MHD approach is appropriate for the macroscopic scales of coronal CSs \citep[][]{snow2018, izquierdo2023}.

The paper is organized as follows. Numerical setup and methods are discussed in Section~\ref{sec:style}.  Our main simulation results are reported and discussed in Section~\ref{sec:floats}. Finally, in Section~\ref{sec:highlight}, we draw our conclusions on the  dynamical and thermal evolution of localized reconnecting current sheets in the presence of different guide-fields in the solar corona.

\section{Numerical setup and methods} \label{sec:style}
Using the PLUTO code \citep{mignone2007pluto}, we setup a numerical model to study the formation, evolution, and physical characteristics of a reconnecting CS in the presence of weak and strong guide-fields, using 2.5-dimensional resistive magnetohydrodynamic (MHD) simulations. We describe below the MHD equations, the details of the numerical domain, the numerical methods, etc.

\subsection{MHD equations}
We solve the following set of non-relativistic resistive MHD equations in two spatial dimensions with all three components of the magnetic and velocity fields retained (i.e.\ a 2.5D system). The system evolves according to the following set of conservation laws, using the PLUTO code \citep{mignone2007pluto}: 
\begin{equation}
\begin{aligned}
\frac{\partial \rho}{\partial t} + \nabla \cdot (\rho \mathbf{v}) & = 0, \\
\frac{\partial \mathbf{m}}{\partial t} + \nabla \cdot \left[ \mathbf{m} \mathbf{v} + \left( p + \frac{B^2}{2} \right) \mathbf{I} - \mathbf{B} \mathbf{B} \right] & = 0 \\
\frac{\partial E_t}{\partial t} + \nabla \cdot \left[ \left(\frac{\rho\mathbf{v}^2}{2} + \rho e + p\right)\mathbf{v}  + c\mathbf{E}\times\mathbf{B}\right] & = 0, \\
\frac{\partial \mathbf{B}}{\partial t} + \nabla \times (c\mathbf{E}) & =0,
\end{aligned}
\label{eq:MHD}
\end{equation}
where $\rho$ represents the mass density of coronal plasma, and $\mathbf{m} = \rho\mathbf{v}$ is the momentum density. Here, $\mathbf{v}$ is the velocity, $p$ is depicted as thermal gas pressure, and $\mathbf{B}$ denotes the magnetic field. The $\mathbf{J}$ is the current density vector and $\eta$ is the magnetic diffusivity. Here, $E_t$ is the total energy density,
\begin{equation}
    E_t = \rho e + \frac{\mathbf{m}^2}{2\rho} + \frac{\mathbf{B}^2}{2}.
\end{equation}
It should be noted that gravitation in the corona is neglected by setting $\mathbf{g}=0$. An ideal equation of state is used to provide closure, $\rho e = \rho e (p, \rho) = \frac{p}{\gamma-1}$. In Eqs.~\eqref{eq:MHD}, the electric field $\mathbf{E}$ (in code-units) is defined by
\begin{equation}
\begin{aligned}
    c\mathbf{E} & = -\mathbf{v}\times\mathbf{B} + \frac{\eta\cdot\mathbf{J}}{c}, \\
    \mathbf{J} & = c\nabla\times\mathbf{B}.
\end{aligned}
\end{equation}

\subsection{Computational domain and normalization factors}
The computational domain for the numerical simulation has a physical size of $x\in [-35, 35]$~Mm in the horizontal direction of the model solar corona, and $y \in [3, 30]$~Mm in the vertical direction of it, further decomposed on a $930 \times 400$ cell uniform grid. This fixes the resolution in the $x$-direction at 75.2~km and in the $y$-direction at 67.5~km. The characteristic scales for length, velocity, and density over which the quantities in Eqs.~\eqref{eq:MHD} are made non-dimensional are $L_0 = 10^8~\mathrm{cm}$, $V_0 = 10^8~\mathrm{cm~s^{-1}}$, and $\rho_0 = 10^{-15}~\mathrm{g~cm^{-3}}$ respectively. The corresponding scales for magnetic field, pressure, temperature, and time are $B_0 = 11.21$~G, $P_0 = 10~\mathrm{dyne~cm^{-2}}$, $T_0 =1.203\times10^8$~K, and $1$~s respectively. The simulation is run for 1800~s (30 minutes) of physical time. For the zero guide-field case before the driver is switched on, the plasma-beta goes from near-infinite at the null-point to values of less than unity at a horizontal distance of $\approx~4$~Mm from the null-point. At the bottom boundary, the plasma-beta is $\approx~0.0014$ at x=0.

\subsection{Initial and boundary conditions}\label{subsec:ibc}
For initial conditions, we assign values typical to the solar corona for density, temperature, and pressure as $5.0\times10^{-15}~\mathrm{g~cm^{-3}}$, $10^6$~K, and $0.3~\mathrm{dyne~cm^{-2}}$ respectively. Pressure (in code units) is computed via the ideal equation of state $p' = \rho'T'/\mu$, where the primed quantities are in code units, as described in the previous section, and $\mu=1.24$. We adopt a uniform resistivity, $\eta = 3.0\times10^{-4}$ in code units, that is typical for solar coronal simulations \citep{Sen2022}. This value is many orders of magnitude higher than the physical resistivity of the solar corona, and is taken to assure that numerical resistivity does not play a significant role in the dynamics. To confirm the dominance of physical resistivity effects over the numerical one, we adopt the method used by \citet{mondal2023reconnection}, in which we perform a comparative study of the thinning phase of CS for chosen zero physical resistivity. The thinning of the CS speeds up in the case of numerical resistivity only, i.e., zero physical resistivity, compared to that for a finite physical resistivity. The increase in $J_{max}$ within the CS also follows the same trend. Hence we can infer that the imposed physical resistivity is higher than the numerical resistivity for the resolution used in this simulation. The Lundquist number ($S_L$) takes values of order 1 in the proximity of the CS, and approaches order 2 as the horizontal distance from the CS increases, for similar values of CS lengths and resistivity to the setup of \citet{syntelis2019cancellation}. In general, these values increase with increasing $B_z$. These values are in agreement with the calculations of \citet{faerder2023comparative}, which computed $S_L$ for the original 2D setup of \citet{syntelis2019cancellation}. $S_L$ is also well below the theoretical limit of plasmoid formation \citep[see][and references therein]{mondal2023reconnection}, which is consistent with the absence of plasmoid like structures in our setup.

We consider the lower $y$-boundary as the base of the corona. This is consistent with the methodology of idealized purely coronal MHD models of flux cancellation  and magnetic field rearrangements \citep{Pontin_2024}, which employ a similar non-stratified coronal box. However, an important distinction here is that the movement of the field lines here is driven by evolving the magnetic field in the ghost  cells at the lower boundary \citep[][]{syntelis2019cancellation}. A future study will consider the effects of stratification on the results.

The initial magnetic field is the sum of fields due to two magnetic sources, a uniform in-plane horizontal field, and an additional out-of-plane guide field: 
\begin{equation}\label{eqn:driver1}
    \mathbf{B} = \frac{F}{\pi}\frac{\mathbf{\hat r}_1}{r_1} - \frac{F}{\pi}\frac{\mathbf{\hat r}_2}{r_2} - b_0\mathbf{\hat x} + b_{g}\mathbf{\hat z}
\end{equation}
where 
\begin{equation}\label{eqn:driver2}
\begin{aligned}
    \mathbf{r}_1 & = (x + d_s)\mathbf{\hat x} + (y-y_0)\mathbf{\hat y},\\
    \mathbf{r}_2 & = (x - d_s)\mathbf{\hat x} + (y-y_0)\mathbf{\hat y}
\end{aligned}
\end{equation}
are the position vectors of the positive and negative magnetic poles (left and right respectively), $d_s=1.8~\mathrm{Mm}$ is the distance of each source from $x = 0$, $y_0 =-0.36~\mathrm{Mm}$ is the vertical location of the charges and $F=1.25\times10^{11}~\mathrm{Mx~cm^{-1}}$ is the magnetic flux associated with each source. $b_0=6~\mathrm{G}$ is the strength of the uniform horizontal field component, and $b_g$ is the strength of the magnetic field component perpendicular to the plane of reconnection, i.e the guide-field. 
The planer magnetic field as given 
by \citet{syntelis2019cancellation} is modified by adding the out-of-plane guide field. Also, the magnetic field parameters are adjusted to make sure that the CS formation takes place completely in the solar corona only, so it may be studied in a specific background condition. We use several values of the guide-field strength in our simulations, namely 0~G, 0.5~G, 1~G, 2~G, 3~G, 4.5~G, 6~G, 9~G, and 12~G to understand its role in the reported formation and subsequent dynamics of the CS. 

At the top ($y=30$) and side ($x=-35$ and $x=35$) boundaries, $\mathbf{v}$ is set to zero, whereas zero-gradient extrapolation is carried out for $\mathbf{B}$, $\rho$, and $p$. At the bottom boundary, $p$ and $\rho$ are fixed to their initial values, $\mathbf{v}$ is set to zero, and the magnetic field is externally driven by converging the two magnetic sources \citep{syntelis2019cancellation} towards $x = 0$ with speed
\begin{equation}
    v_{m}(t) = \frac{1}{2}v_{max}\left(\tanh\frac{t-t_0}{w} + 1\right),
    \label{eq:vt}
\end{equation}
where $v_{max}=1~\mathrm{km~s^{-1}}$, $t_0=10.1~\mathrm{minutes}$ and $w=1.4~\mathrm{minutes}$.
Physically, $t_0$ denotes the time after which the magnetic sources commence their displacement, and in this way, at time $t>t_0$, each magnetic source has moved a distance
\begin{equation}
    x_{m}(t) = v_{max}\frac{w}{2}\left[\ln{\cosh{\frac{t-t_0}{w}}} - \ln{\cosh{\frac{t_0}{w}}}\right] + \frac{1}{2}v_{max}t
    \label{eq:xt}
\end{equation}
toward the $y$-axis. By replacing $d_s$ in Eqs.~\eqref{eqn:driver1} and \eqref{eqn:driver2} with $d_s - x_{m}(t)$, the simulation is driven solely by varying the value of $\mathbf{B}$ in the ghost cells at the bottom boundary $y=3$~Mm. Hence, the formation of and reconnection within the CS is driven by the response of the corona and coronal magnetic field to changing magnetic fields at $y = 3$~Mm, which essentially mimics the convergence of two magnetic sources placed below the model corona. The evolving magnetic configuration induces highly localized self-consistent plasma flows in the interior, particularly at the footpoints of the converging flux patches. Setting $\mathbf{v}=0$ and fixed density and pressure essentially prevents artificial flows through the lower boundary after the driver is switched on. 

\subsection{Numerical methods}
Spatial interpolation is performed with a third-order weighed non-oscillatory reconstruction scheme \citep[WENO3;][]{yamaleev2009weno}. A third-order Runge--Kutta time-stepping is used, and we set the Courant--Friedrichs--Lewy limit to 0.3. We use a Harten--Lax--van Leer (HLL) Riemann solver \citep[][]{harten1983upstream,janhunen2000positive} for flux computation at cell interfaces. The solenoidal constraint is maintained using the constrained-transport method \citep[][]{evans1988simulation,balsara1999staggered} with a staggered representation of the magnetic field using both face- and cell-centered field values. In this way, the solenoidal condition is maintained to machine accuracy. Due to the use of a staggered grid, we define the initial planar magnetic field (in X-Y plane) in terms of a vector potential given by
\begin{equation}
    \begin{aligned}
        A_x &= 0,\\
        A_y &= 0,\\
        A_z &= \frac{F}{\pi}\tan^{-1}\left(\frac{x-d_s}{y-y_0}\right) - \frac{F}{\pi}\tan^{-1}\left(\frac{x+d_s}{y-y_0}\right) - b_0y.
    \end{aligned}
\end{equation}


\section{Results and discussion} \label{sec:floats}
\begin{figure*}[h]
\includegraphics[width=0.9\linewidth]{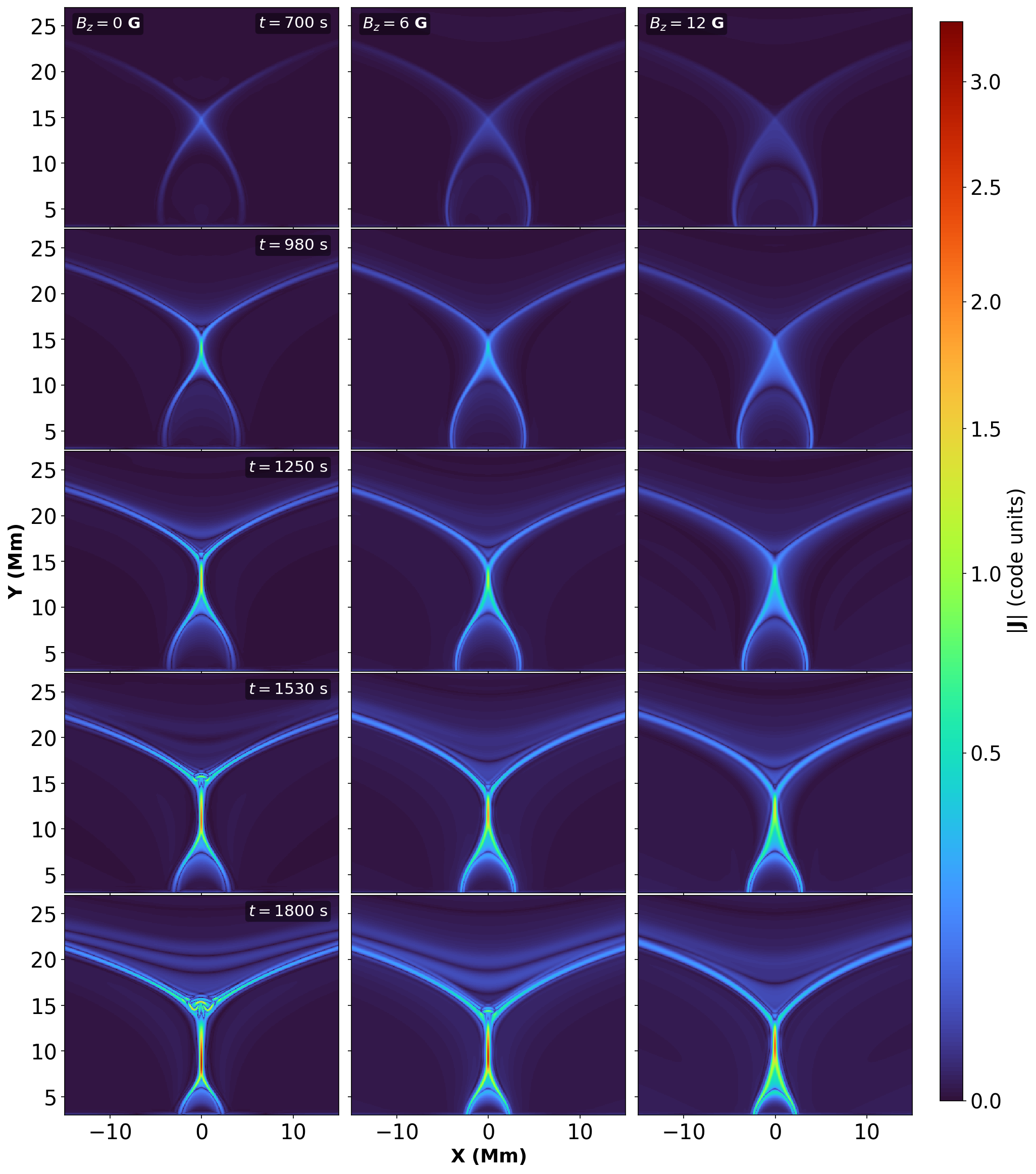}
\caption{Snapshots of the magnitude of the current density vector, $\mathbf{J} = J_x\hat{x} + J_y\hat{y} + J_z\hat{z}$ at times 700, 980, 1250, 1530 and 1800~s (top to bottom) for the magnetic-field component perpendicular to the plane of reconnection, i.e., the guide-field ($B_z$) of 0, 6 and 12~G (left to right). The figure captures the evolution of the developing CS structure as the magnetic null collapses in response to the converging motion of the opposite magnetic flux sources, leading to the formation and subsequent reconnection of the CS. At a given time, the region of high current density is less pronounced for higher $B_z$. An animation covering the entire dynamics of the CS for all these three different guide-fields is available in the online version with a real time duration of 12~s The associated movie is available online.
\label{fig:mosaic_J_vec}}
\end{figure*}

We perform a numerical simulation of the formation, dynamical evolution, and heating of a reconnecting CS due to converging motion of the field lines. The converging motion of magnetic polarities as a moving magnetic feature (MMF) at the coronal base may lead to flux cancellation or alternatively rearrange the field lines in the overlying atmosphere \citep[e.g.,][]{priest2018cancellation,syntelis2019cancellation,2025ApJ...989..222M}. In the present study, we only mimic the motion of the sources at the bottom boundary of the simulated corona that further leads such rearrangements in the localized corona around the null. Field-line movements subsequently lead to the collapse of the null, and rearrangements further cause reconnection within the evolving CS in the presence of different guide-fields (see Figure~\ref{fig:mosaic_J_vec}). It can be inferred from Equations~\eqref{eq:vt} and \eqref{eq:xt} that for time $t<t_0$ the speed of convergence of the magnetic sources is negligible, hence the distance between the sources does not change. Therefore, the system remains in equilibrium until $t_0$ with the null points remaining unperturbed and stationary at their initial height in the overlying corona. 
Hence, reconnection is not taking place for $t<t_0$. Therefore, negligible nonzero variations observed in the current density, $\mathbf{J}$, before $t_0$ are numerical artifacts possibly due to the proximity of the magnetic flux sources to the line-tied lower boundary and hence not considered in any analysis presented here. 

Since $B_z$ can now influence in-plane dynamics through the pressure-gradient term, it is important to estimate whether the relative contribution between thermal and guide-field pressure changes appreciably throughout a simulation. We find that this is not the case, and the relative contribution stays within the same order, which is determined by the initially set value of $B_z$. For more details, see Appendix~\ref{app:bzvsp}.

We run the simulation for a total of 1800 s.  
All relevant quantities are analyzed from $t = 700$~s, i.e., 94~s after $t_0$. Essentially, we try to ensure that  
the values of $|\mathbf{J}|$ around the collapsed null-point are greater by several orders of magnitude than the values of the current density developed during $t<t_0$ for different values of the guide magnetic field. Therefore, $t$=700~s is taken as the initial time for the physical collapse of the null, evolution of CS, and its magnetic reconnection process for further analysis. From 700~s onward, current starts to accumulate gradually at the null point and at the separatrices for all values of the guiding magnetic field (see Figure~\ref{fig:mosaic_J_vec}).   
We also conduct a test run with $t_0=0$~minutes for $B_z =$ 12~G, and find that the current starts to accumulate at $t\approx140$~s, and this time interval decreases for decreasing $B_z$. Therefore, it appears that the collapse of oppositely directed field lines within higher plasma-beta regions begins around 100~s after initiation of the boundary driving.

Although the provided maximum velocity of the flux sources (which are outside the computational domain) is assigned to be $1~\mathrm{km~s^{-1}}$, the observable and measurable speed of convergence of the representative sources, that is, the opposite magnetic polarity peaks at the lower boundary, is $v_{0,m}=0.3$--$0.4~\mathrm{km~s^{-1}}$. Such an apparently low value of the convergence speed is reported by \citet{hou2021formation} in their observations from the HMI vector magnetogram. In the absence of a stratified atmosphere, we essentially consider the convergence of opposite polarities at the lower boundary 
as converging magnetic connectivities in the localized corona, which are  a likely response of those movements at the base of the corona where the field lines are re-arranged slowly. At the lower y boundary, the magnitude of the maximum magnetic field is found to be $\approx80~\mathrm{G}$, which is consistent with the values observed in the active solar corona \citep[e.g.,][]{2Alissandrakis2021cormag,brooks2021cormag}. The reported value is measured at time $t<t_0$, prior to the driving of magnetic field lines. In the present work, our main aim is to investigate the self-consistent morphology,  evolution, and heating of the CS in the presence of a wide range of guide-fields with respect to the overlying field strength $b_0$, which we do so in the following subsections.

\subsection{Morphological differences in the formation and evolution of CS in presence of different guide-Fields}
In the absence of any guide-field component, the field lines connecting opposite polarities are exactly opposite of the overlying horizontal field lines at certain locations, and this gives rise to a ``perfect'' magnetic null with magnetic-field magnitude being close to zero. However, in the cases with nonzero guide-field values, the resultant magnetic field will not be close to zero even at the null location, thanks to the non-zero contribution of a guide-field component. Therefore, the plasma-beta values at the null location for $B_z$ of 0.5, 1, 2, 3, 4.5, 6, 9 and 12~G are 33.59, 8.42, 2.11, 0.94, 0.42, 0.23, 0.10 and 0.06 respectively. For the case of 0~G, a very high plasma-beta is measured at the null point. These numbers indicate that the plasma-beta decreases from the high plasma-beta regime at the true magnetic null ($B_z~=~0$~G) to lower values at the in-plane null for finite guide fields.

The current density begins to accumulate very slightly in the null region and separatrices within 606-700 s (see the first row of Figure~\ref{fig:mosaic_J_vec}), for all cases. At t = 700 s, the magnitude of maximum current density are  
32.14, 31.75, 30.66, 27.15, 23.26, 18.22, 15.03, 10.83 and 8.46 statA~cm$^{-2}$
for increasing values of $B_z$.  The highest peak is observed for $B_z=0$~G.

By 980 s, the null point has visibly collapsed, and sheet elongation has been initiated for each case (see Figure~\ref{fig:mosaic_J_vec}). More importantly, the formation of the CS and specifically high current density regions within the CS are less pronounced at higher values of guide-fields (see Figure~\ref{fig:mosaic_J_vec}). This clearly suggests that the presence of different guide-fields are playing a significant role in the formation and evolution of the CS. The region near the lower boundary shows arc-shaped density and pressure enhancements in the lower arcade over the polarity-inversion-line (PIL), consistent with the reconnection flow towards the magnetic flux sources delivering mass to a magnetically closed region. High density arc-like regions also correspond to similarly structured high pressure regions. However, the simulation develops self-regulating strong field-aligned drainage flows at the footpoints of the arc, confirmed to be physical features by their presence in high-resolution runs. Hence, we further perform a detailed analysis of the reconnection within the dynamical CS and its physical outcomes in the following subsections.

\begin{figure*}[ht!]
\centering
\includegraphics[width=\linewidth]{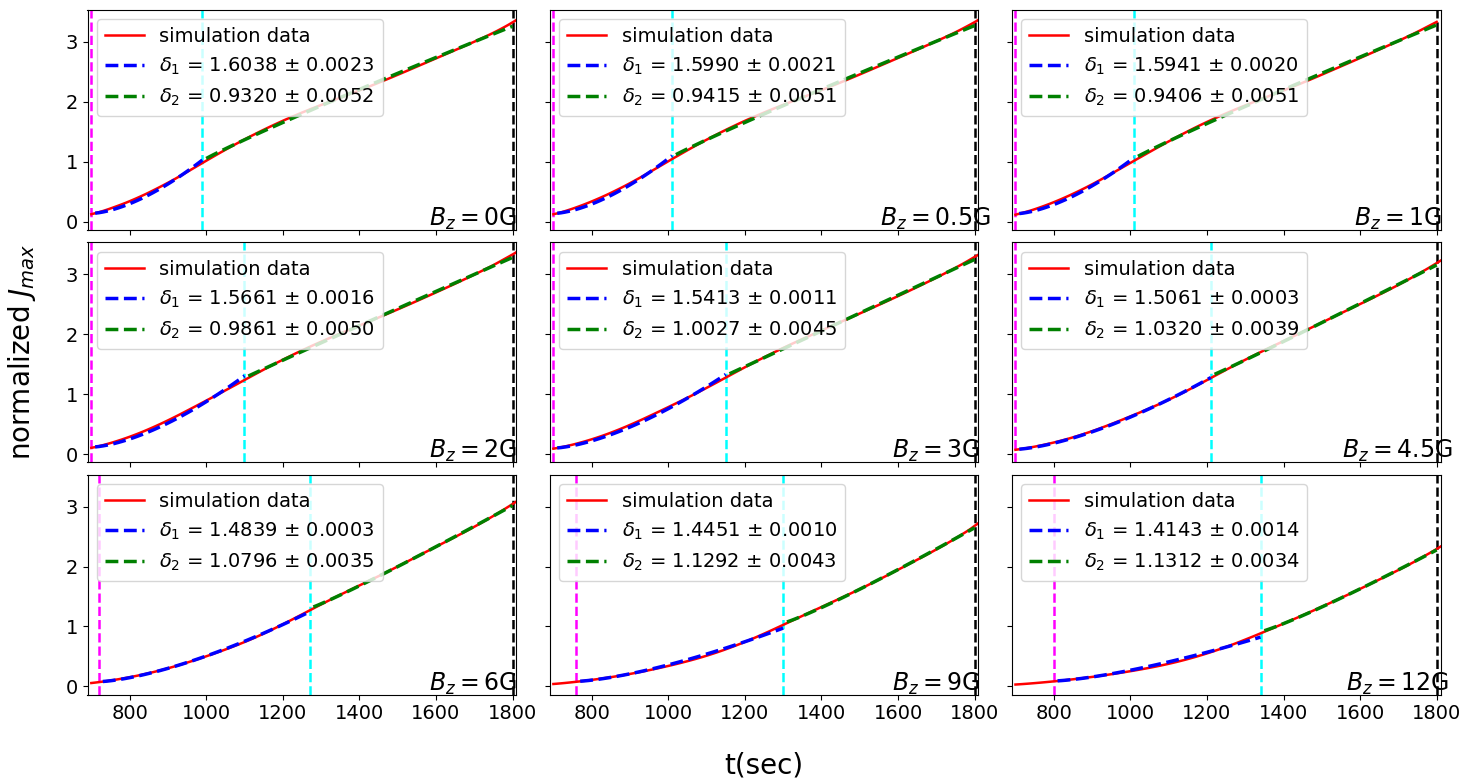}
\caption{Power-law fitting of the temporal profile of the maximum magnitude of the current density within the computational domain for multiple values of the guide-field $B_z$. For optimal fitting of the early-stage profile (within the time window indicated by the vertical magenta and cyan dashed lines) with exponent $\delta_{1}$, the proportionality constant ($D$) is kept as $10^{-4}$ for all $B_{z}$. To accommodate the transition of the exponent $\delta_2$ from less to a greater than unity while fitting later stage profiles, the proportionality parameter $D$ and exponent $\delta_2$ are fitted simultaneously after $\theta_{2}$~s, denoted via dashed cyan lines.  
The magenta line corresponds to the start of the time window for studying early-stage reconnection, i.e., $\theta_1=$700, 700, 700, 700, 700, 700, 720, 760, and 800~s respectively for increasing strength of the guide-field, whereas cyan dashed lines indicate the end of that time window, i.e., $\theta_2=$990, 1010, 1010, 1100, 1150, 1210, 1270, 1300, and 1340~s respectively for increasing strength of the guide-field. Black dashed lines correspond to $\theta_3=1800$~s, which is the end point of studying late-stage reconnection.
}
\label{fig:etajmax}
\end{figure*}

\subsection{Temporal evolution of reconnection rate from 700 to 1800 s} \label{sec:3.2}
Computing the reconnection rate is an important step to understand the dynamics of ongoing reconnection within the CS. Following \citet{yokoyama2001flare} and \citet{mondal2024CS}, we consider $\eta J_{max}$ as a proxy for the reconnection rate, where $\eta$ is the uniform, time-independent resistivity in the present simulations. Hence, we analyze the temporal evolution of $J_{max}$ for all values of $B_z$, as the reconnection rate would then be directly proportional to $J_{max}$, and the fitted exponents would follow the same trend as found here, albeit with different values. We calculate it by measuring the maximum magnitude of the current density $\mathbf{J}$, within the computational domain, excluding the values of $\mathbf{J}$ at the boundaries. Within the time window under consideration, $J_{max}$ and hence $\eta J_{max}$ increases monotonically with time for all values of $B_z$ (see Figure~\ref{fig:etajmax}). 

To better quantify the dependence of the temporal variation of the reconnection rate on different guide-field strengths, we fit the profiles of $J_{max}$ for each $B_z$ with a power law of the form
\begin{equation}\label{eqn:delta}
    J_{max}(t) = D(t - \theta)^\delta + J_{max}(\theta),
\end{equation}
where $\delta$ is the exponent of best fit for the data set, $D$ is a proportionality factor, and $J_{max}(\theta)$ is the value of $J_{max}$ at $t=\theta$~s, so that the fitted function and datapoint coincide at that particular time. Since we have a high variance of guide-field values starting from very low (0.5~G) to values as high as 12~G, including the strictly 2D case of 0~G, we need to ensure that we are fitting the right time window in all the cases before drawing any comparison between cases. The measured profiles of $J_{max}$ are such that they cannot be fitted with a single fitting curve. So, to determine the time interval, we rely on the time derivative of the $J_{max}$ profile, such that the derivative should show a rapid increase at the start of the window and then a decrease at the end of the window. We denote the start and end of such a time window as $\theta_{1}$ and $\theta_{2}$ in Figure~\ref{fig:etajmax} (see Appendix~\ref{sec:app1} for details on $\theta_1,~\theta_2$ calculation). Visual inspections shows that there may be a different slope with which $J_{max}$ is changing at a later phase. Hence, we fit that time interval starting from t = $\theta_{2}$~s up to $\theta_3=1800$~s separately (indicated as black dashed lines in the same Figure). Figure~\ref{fig:etajmax} shows that the break point (vertical cyan line) of the two phases of $J_{max}$ is shifting to the right with a higher value of $B_z$. 

For fitting within $\theta_{1}$ to $\theta_{2}$~s using Equation~\eqref{eqn:delta}, $D$ is taken to be $10^{-4}$ for all values of $B_{z}$ as determined by order-of-magnitude comparison between the time scale and the concerned variable scale and $\theta = \theta_1$ is considered. Then the exponents of best fit within such case-specific time windows are listed in Figure~\ref{fig:etajmax} as $\delta_1$.
 
At a later phase after $t=\theta_{2}$~s, i.e., for fitting within $\theta_{2}$ to $\theta_{3}$~s, the parameter $D$ needs to be case specific rather than fixed at a certain value for best fitting, and $\theta_2$ is considered in place of $\theta$ in Equation~\eqref{eqn:delta}. So, the values of $D$ are $(4.365 \pm 0.1435) \times 10^{-3}, (4.093 \pm 0.1326) \times 10^{-3}, (4.191 \pm 0.1355) \times 10^{-3}, (3.142 \pm 0.09813) \times 10^{-3}, (2.926 \pm 0.08125) \times 10^{-3}, (2.558 \pm 0.06060) \times 10^{-3}, (1.964 \pm 0.04059) \times 10^{-3}, (1.424 \pm 0.03662) \times 10^{-3}, (1.327 \pm 0.02674) \times 10^{-3}$ for best fitting in cases with increasing $B_{z}$. More importantly, the values of the exponent, $\delta_{2}$ are listed in Figure~\ref{fig:etajmax}.

Even though for small guide-field values of 0 G, 0.5 G, and 1 G the values  of the fitted exponents are nearly equal to each other in both phases within the fitting uncertainties, small differences are still present. In the early stage, $\delta_1$ shows a decreasing trend with increasing guide-field strength, indicating that the temporal growth of $J_{max}$, and hence of the reconnection-rate proxy $\eta J_{max}$, becomes slower for stronger guide fields. This suggests that a stronger guide field slows the early-stage development of reconnection. Consistent with this trend, the break point $\theta_2$ is delayed to later times with increasing $B_z$. In contrast, $\delta_2$ shows an increasing trend with increasing guide-field strength in the time interval $\theta_2~-~\theta_3$, indicating a faster temporal growth of $J_{max}$ at the later stage with increasing guide-field values.

As far as observational perspectives of magnetic reconnection are concerned, $J_{max}$ or $\eta J_{max}$ are not directly observable. Rather, the aspect ratio of the CS, bidirectional outflows, and temperature enhancements in the form of radiation-intensity enhancements are more detectable evidence of underlying reconnection processes. Therefore, we next analyze other measurable features, such as the aspect ratio, maximum outflow velocities, and maximum temperature within the simulated reconnecting CS, to assess whether a similar trend persists in these observables.

\begin{figure*}[ht!]
\centering
\includegraphics[width=\linewidth]{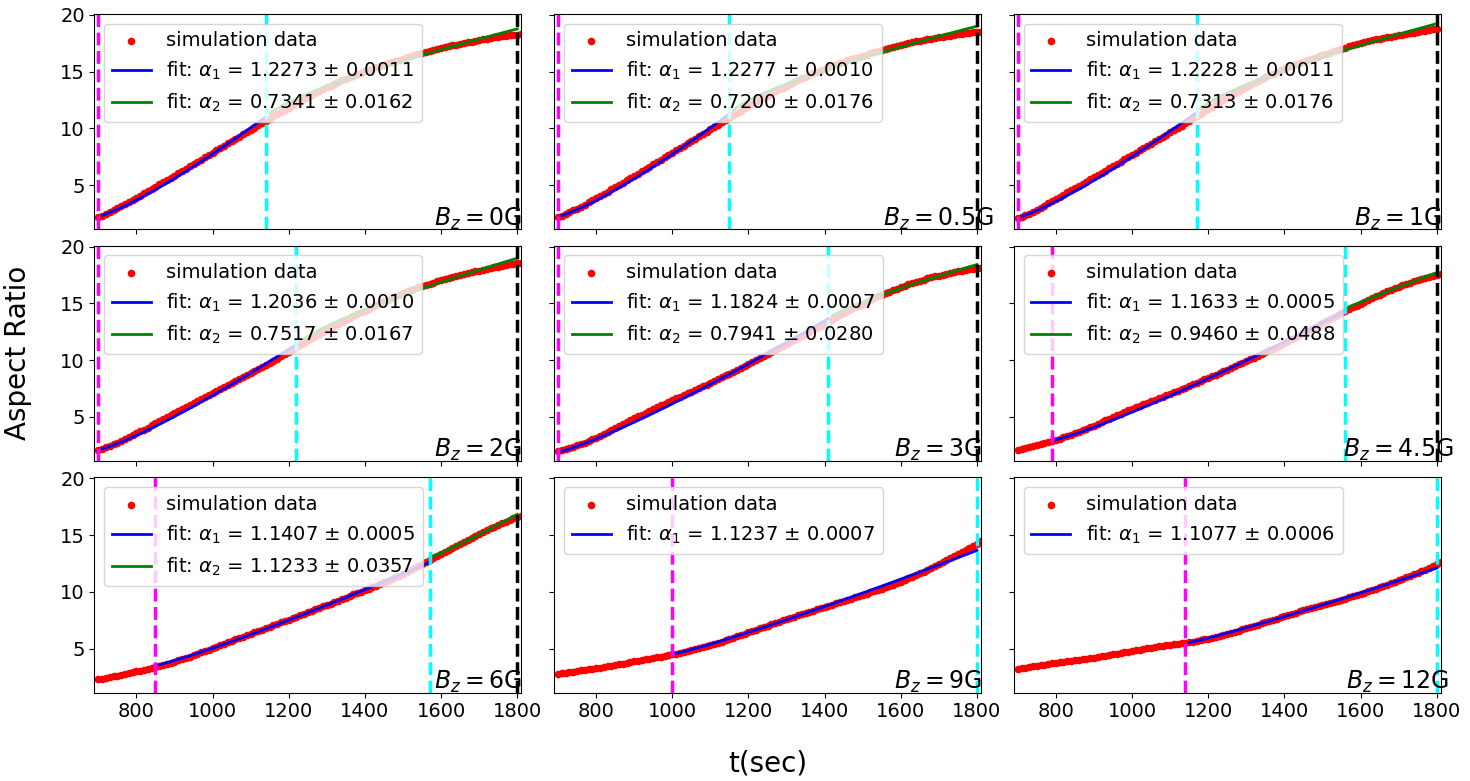}
\caption{Power-law fitting of the temporal profile of the CS aspect ratio for multiple values of the guide-field $B_z$. Here, we fitted profiles within time windows denoted by the vertical magenta and cyan ($\phi_1-\phi_2$~s) and cyan and black ($\phi_2-\phi_3$~s) dashed lines separately. As the temporal derivative of the aspect ratio changes from a monotonically increasing to a decreasing profile around the vertical cyan line, we have fitted with different power laws in two regions with exponents, respectively, $\alpha_1$ and $\alpha_2$. The free parameter of multiplication $A$ is set to $5.0 \times 10^{-3}$ for optimal fitting of $\alpha_1$. For $\alpha_2$, the proportionality parameter $D$ is fitted simultaneously with $\alpha_2$ to accommodate its transition from lesser to greater than unity. For lower values of the guide-field (0--3~G), the aspect ratio peaks, and conversely it does not reach higher values for the larger $B_z$. $\phi_1$ corresponds to times 700, 700, 700, 700, 700, 790, 850, 1000, and 1140~s respectively (vertical magenta dashed line) for increasing strength of the guide-field. $\phi_2$ corresponds to times 1140, 1150, 1170, 1220, 1410, 1560, 1570, 1800, and 1800~s respectively (vertical cyan dashed line) for increasing strength of the guide-field. 
\label{fig:aspect_ratio_curve_fit}
}
\end{figure*}

\begin{figure*}[ht!]
\centering
\includegraphics[width=0.85\linewidth]{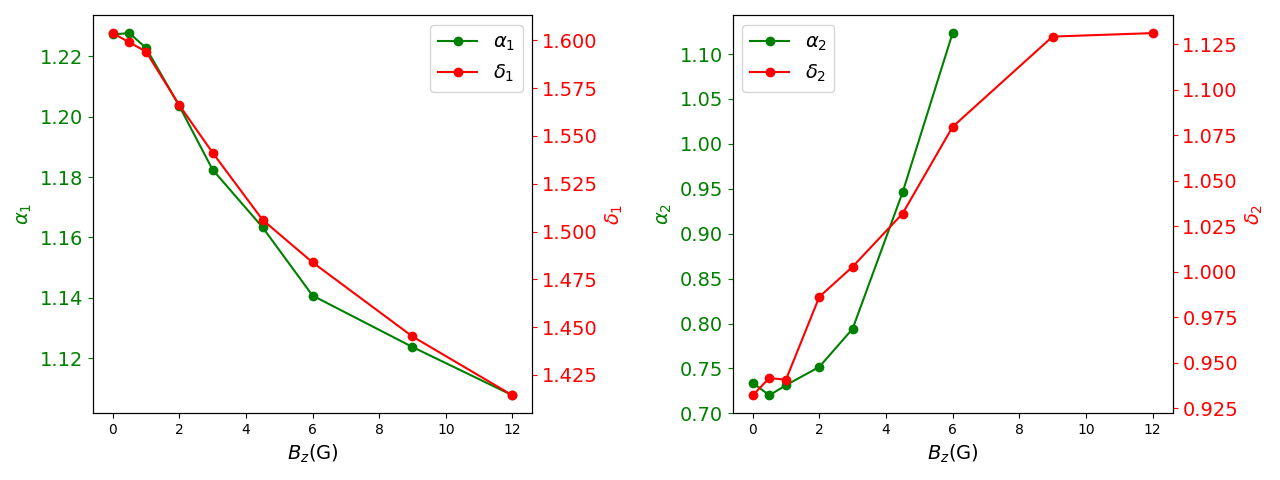}
\caption{Left panel: Dependency of the exponents $\alpha_1$ and $\delta_1$ of $J_{max}$ and aspect ratio of the CS, respectively, on guide-field $B_z$, as described in Figures~\ref{fig:aspect_ratio_curve_fit} and \ref{fig:etajmax}. Right panel: Evolution of exponents $\alpha_2$ and $\delta_2$ with respect to $B_z$ for the phase bounded by the cyan-black vertical dotted lines in $ J_{max}$ (and hence, reconnection rate) and aspect ratio, respectively. The scales and plots of $\alpha$ and $\delta$ exponents are marked in green and red respectively for both figures.
\label{fig:alpha_delta}
}
\end{figure*}

\subsection{Temporal evolution of the aspect ratio of the CS from 700 to 1800~s}\label{sec:aspectratio}
We verify that even though there are nonzero values of $B_z$ in most of the cases, values of $J_{x}$ and $J_{y}$ are negligible (at least one order of magnitude smaller) in comparison to the value of $J_{z}$ once CS formation starts in the interval 700--1800~s.
So, to define the length ($L$) and width ($W$) of the CS, we obtain the values of $|J_z|$ along two cuts, say, segments AB and CD parallel to the $y$- and $x$-axis respectively. In this way, we estimate $L/W$ as the aspect ratio of the CS. Since the CS does not move along the horizontal direction, segment AB is fixed at $x = 0$~Mm. On the other hand, the CS has a downward motion at later stages. Hence, the location of segment CD has to be instantaneous rather than fixed. To determine its location, we consider the $y$-position of maximum $|J_z|$, and segments AB and CD intersect there. We then calculate the distances along $y$ and $x$ where $|J_z|$ drops to 50\% of the instantaneous maximum from the intersection point along AB and CD. Such distances are considered as the half-length and half-width respectively, and therefore used to calculate $L$, $W$, and $L/W$ as the aspect ratio.

Figure~\ref{fig:aspect_ratio_curve_fit} shows the variation of the aspect ratio with time, for all values of $B_z$ considered. Similarly to $J_{max}$, the aspect ratio evolves in two phases over time, indicated by the region of magenta-cyan and cyan-black dotted lines. In these phases, we fitted power-law curves with exponents, respectively, $\alpha_1$ and $\alpha_2$ using the fitting function 
\begin{equation}\label{eqn:alpha}
    \frac{L(t)}{W(t)} = A(t - \phi)^\alpha + \frac{L(\phi)}{W(\phi)},
\end{equation}
where $\phi$ is taken to be $\phi_1$ and $\phi_2$ for fitting within intervals of $\phi_1$--$\phi_2$ and $\phi_2$--$\phi_3$ respectively. For details on the calculation of $\phi_1,$ $\phi_2$, and $\phi_3$, see Appendix~\ref{sec:app1}. The time intervals of these two phases  are not the same as the intervals $\theta_1$--$\theta_2$ and $\theta_2$--$\theta_3$ for $J_{max}$ discussed before, but the trend of delaying the break point in time with higher $B_z$ is similar to that displayed by $J_{max}$ and reconnection rate, which indicates a possible correlation of both quantities with the initially applied guide-field. See Appendix~\ref{sec:app2} for more details. 

The corresponding exponents of the best power-law fit $\alpha_1$ for increasing $B_z$ are 
indicated in Figure~\ref{fig:aspect_ratio_curve_fit}. The exponents are obtained by fitting data between the magenta ($t =\phi_1$~s) and cyan ($t =\phi_2$~s) dashed lines. These values indicate that the exponent is decreasing with an increase in the value of $B_{z}$, which means that thinning and elongation of the CS are halted to some degree in the presence of strong guide-fields. Similarly, the exponents $\alpha_2$ for the best power-law fit in the region bounded by cyan-black lines (time interval $\phi_2-\phi_3$~s) are also indicated in Figure \ref{fig:aspect_ratio_curve_fit}.
 
However, for $\alpha_2$, the proportionality constant $A$ is fitted simultaneously with $\alpha_2$, to account for its observed behavior, and it assumes the following values: $(6.657 \pm 0.6567) \times 10^{-2}, (7.341 \pm 0.7880) \times 10^{-2}, (7.040 \pm 0.7519) \times 10^{-2}, (6.477 \pm 0.6474) \times 10^{-2}, (4.194 \pm 0.6579) \times 10^{-2}, (1.777 \pm 0.4477) \times 10^{-2}, (0.836 \pm 0.1538) \times 10^{-2}$. It is to be noted that this power-law fitting of the aspect-ratio profile at a later stage of reconnection has been performed only up to a guide-field of 6~G, as for 9 and 12~G the break point shifts to a time beyond our final simulation time of 1800~s. To confirm this, we re-run the simulations for 9~G and 12~G till 2400~s and observe that the data-window for calculating $\alpha_2$, as described in Appendix~\ref{sec:app1} shifts beyond the 1800~s mark for guide-field strengths of 9~G and 12~G. 

With this analysis, we can correlate the exponents with which $J_{max}$ and aspect ratio evolve with time, i.e., $\delta$ and $\alpha$, to find whether there is any dependency of the entire reconnection process on the guide-field in terms of observables, e.g.\ the aspect ratio, as well as non-observables, e.g.\ $J_{max}$. In Figure~\ref{fig:alpha_delta}, we show the profile of $\alpha_1$, $\delta_1$, $\alpha_2$, and $\delta_2$ vs.\ $B_z$ as estimated within the time windows mentioned above. The decreasing and increasing trend of $J_{max}$ and aspect ratio at early and later phases, respectively, with increasing $B_{z}$ suggests a mutual correlation between how the evolution of  $J_{max}$ and aspect ratio ($L/W$) with time is affected by $B_{z}$. The increasing trend in $\alpha_2$ and $\delta_2$ suggests that the reconnection-inhibiting effect of the guide-field is reducing in strength, possibly due to a reduction in the strength of the average guide-field strength itself and continuously increasing inflows (see Appendix~\ref{sec:app2}).

\begin{figure*}[ht!]
\centering
\includegraphics[width=\linewidth]{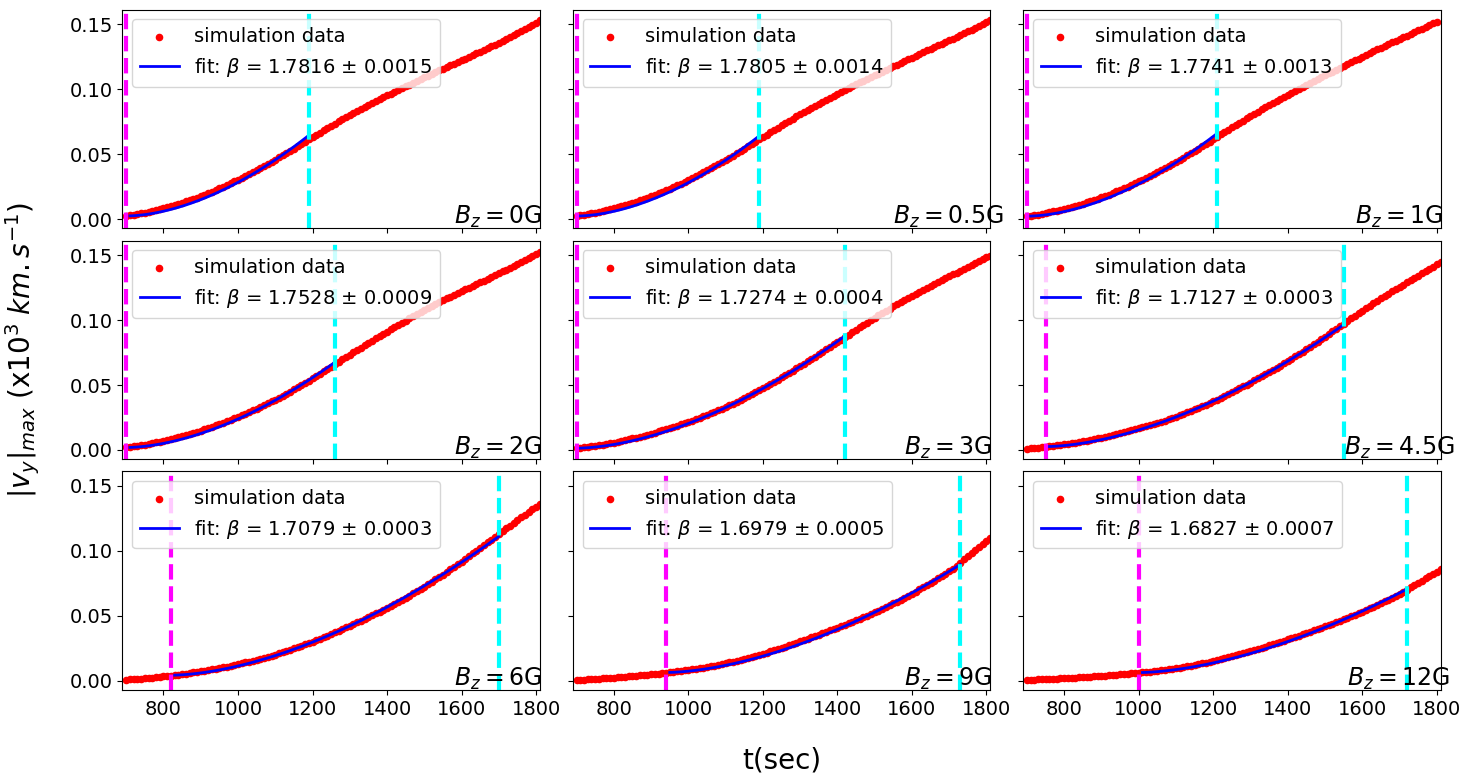}
\caption{Power-law fitting of the temporal variation of maximum absolute upflow velocity, i.e., $|v_{y}|_{max}$, along the reconnecting CS for multiple values of the guide-field $B_z$. The power-law exponent $\beta$ quantifies a phase of increasing growth of $|v_y|_{max}$ within a time window bounded via dashed magenta and cyan lines. The free parameter of multiplication $C$ is set to $1.0 \times 10^{-6}$, for optimal fitting of all data sets. Magenta dashed lines denote the start of the time window of power-law fitting, $\psi_{1}=$700, 700, 700, 700, 700, 750, 820, 950, and 1000~s respectively for increasing strength of the guide-field. cyan dashed lines show the end of the time window of the power law fitting at $\psi_{2}=$1190, 1190, 1210, 1260, 1420, 1550, 1700, 1730, and 1720~s for increasing values of the guide-field. After $\psi_{2}$~s, we still observe a growth of $|v_y|_{max}$ but with a lower rate, which we exclude while fitting.
\label{fig:max_vy}
}
\end{figure*}

\subsection{Temporal evolution of bidirectional outflows from 700 to 1800~s}\label{sec:vymax}
While reconnection is taking place, magnetic energy is converted to kinetic energy as well as heat. Hence, bidirectional CS outflows are one of the observational consequences of the reconnection process. Also, henceforth, we refer to upward and downward directed plasma outflows from the top and bottom halves of the CS as upflows and downflows respectively. The generation of more kinetic energy at a faster rate may lead to higher outflow velocities. Therefore, higher outflow velocities can be safely attributed to faster reconnection \citep[e.g.,][]{mondal2025CS2}. Hence, we investigate the presence of any scaling relation between maximum absolute upflow velocity ($|v_{y}|_{max}$ in the present case) and the value of the guide-field. The maximum upflow velocity $|v_{y}|_{max}$ is calculated from within the region of $L(t) \times W(t)~\mathrm{Mm}^{2}$ centered around the instantaneous location of the maximum current density, where $L(t)$ and $W(t)$ are the time-dependent length and height of the CS as described in the previous section. 

As expected, at around 700~s the upflow velocity is almost zero in all nine simulations, since reconnection is yet to start or has just started. At 1800~s, the maximum upflow velocity along the CS reaches approximately 150, 151, 152, 149, 147, 144, 135, 106, 82~$\mathrm{km~s^{-1}}$ 
for increasing guide-field strength (see Figure~\ref{fig:max_vy}). Hence, we find that upflow speeds are decreasing for stronger guide-fields as a consequence of a slowdown of the underlying magnetic reconnection. To quantify such trends, we perform a power-law fitting of the growth of the upflow velocity with time within the magenta ($t =\psi_1$~s) and black ($t =\psi_2$~s) dashed lines (see Appendix~\ref{sec:app1} for discussion on such choices of time window) using the following fitting function
\begin{equation}\label{eqn:beta}
    |v_{y}|_{max}(t) = C(t - \psi_1)^\beta + |v_{y}|_{max}(\psi_1),
\end{equation}
where t is taken to be between $\psi_1$ and $\psi_2$. The corresponding exponents of the best power-law fit $\beta$ are indicated in Figure~\ref{fig:max_vy}.

For weak guide-fields, such as 0~G and 0.5~G, the values of $\beta$ overlap almost completely with each other within fitting uncertainty. Since uncertainties in those three cases are very close to each other, such overlaps indicate that for weak guide-fields the upflow speed is not strongly affected by the absence or presence of a guide-field. However, for much higher values of the guide-field, it is clear that $\beta$ is decreasing with increasing $B_{z}$. This suggests that the presence of strong guide-fields strongly affects the upflow speed, and hence the kinetic energy generated via magnetic reconnection. 
However, this setup possesses an asymmetry between the upflows and downflows emerging from the CS. Upon tracking the maximum downflow speed from the lower half of the CS (i.e., the region under the point of maximum $\mathbf{J}$ at that time), we observe a trend similar to the upflow velocity, with a few differences. For a given $B_z$, the growth rate of the downflow velocity is invariably smaller than the growth rate of upflow velocity. With increasing $B_z$, the growth exponent of the downflow speed drops, which is similar to the effect of increasing $B_z$ on upflows. 
 
The relative closeness of the lower boundary to the CS plays a minimal role in the asymmetry between the observed upflows and downflows since there is no deceleration observed in the downflow values throughout the simulation time. Also, there are no emerging upward flows or significant gradients in the region below the CS that could impede the downflow of plasma from the CS.

\begin{figure*}[ht!]
\centering
\includegraphics[width=\linewidth]{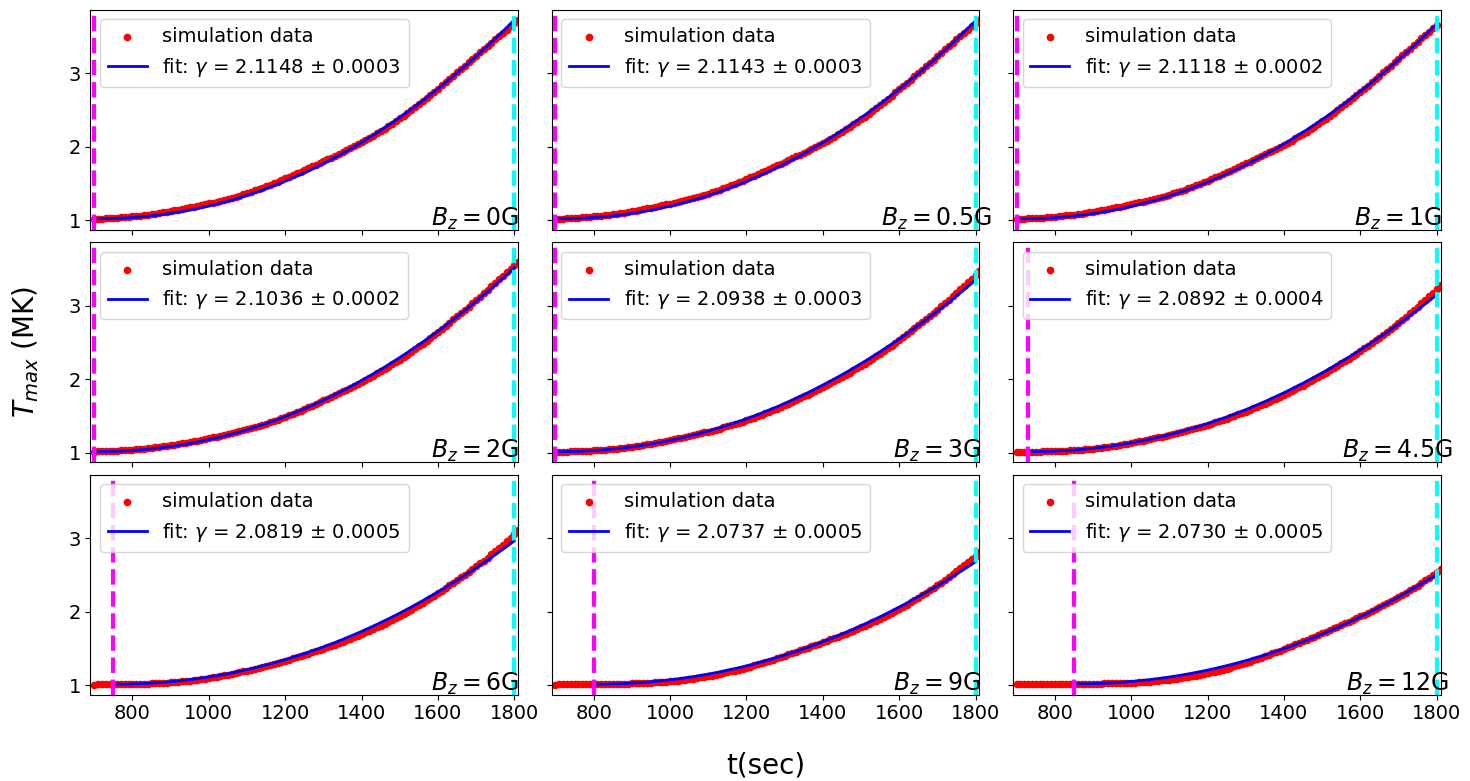}
\caption{Power law fitting of the variation in maximum temperature with time within the CS for multiple values of the guide-field $B_z$. The free parameter of multiplication $E$ is set to $1.0 \times 10^{-6}$. Temperature enhancement is visibly delayed and/or slowed down for increasing values of $B_z$, which indicates a trend towards less heating and slower reconnection. This is considered while choosing a time interval for optimal fitting. $\chi_1$ corresponds to times 700, 700, 700, 700, 700, 730, 750, 800, and 850~s respectively (vertical magenta dashed line) for increasing strength of the guide-field. $\chi_2$ corresponds to a time of 1800~s (vertical cyan dashed line) for all values of the guide-field.
\label{fig:max_temp}
}
\end{figure*}

\subsection{Temporal evolution of maximum temperature from 700 to 1800~s}\label{sec:tmax}
In imaging observations through various space observatories, high intensity enhancements at localized regions are mostly attributed to underlying occurrence of energy release and heating via magnetic reconnection. Even though we cannot probe temperature and density of dynamic plasmas directly from imaging observations, specific techniques such as Differential Emission Measure (DEM) analysis can be used to deduce information about density and temperature of emitting plasma columns. In contrast, we have direct measurement of temperature and density in numerical simulations. Hence, we investigate the temperature evolution in the aforementioned nine simulation cases over time.

We compute the maximum temperature within the rectangular domain $L(t) \times W(t)~\mathrm{Mm}^{2}$ centered around the instantaneous location of the maximum current density as before. We find that maximum temperature values are around 1~MK in all nine cases at about 700~s, i.e., no heating is taking place yet. For lower values of $B_{z}$, between 0~G and 4.5~G, temperature starts increasing around 700~s (see Figure~\ref{fig:max_temp}). However, for even higher guide-fields, temperature enhancement commences at later times, i.e., roughly around 700, 800, and after 850~s respectively for $B_{z}$ of 6~G, 9~G and 12~G (see Figure~\ref{fig:max_temp}). At 1800~s, the maximum temperature reaches around 3.68, 3.685, 3.66, 3.55, 3.42, 3.23, 3.06, 2.76, and 2.54~MK while $B_{z}$ increases from 0~G to 12~G (actual values are reported above). 

In accordance with the analyses of previous sections, we perform power-law fitting of the temporal dependence of the maximum temperature inside the CS within magenta ($t=\chi_1$~s) and cyan ($t =\chi_2$~s) dashed lines (see Appendix~\ref{sec:app1} for a discussion on the choice of time windows) using the fitting function
\begin{equation}\label{eqn:gamma}
    T_{max}(t) = E(t - \chi_1)^\gamma + T_{max}(\chi_1),
\end{equation}
where t is taken between $\chi_1$ and $\chi_2$ in Figure~\ref{fig:max_temp}. 
The corresponding exponents of the best power-law fit $\gamma$, in order of increasing $B_z$, are also indicated with their respective guide-field case in Figure~\ref{fig:max_temp}.
Hence, it appears that the presence of higher guide-fields results in lower rates of enhancement of temperature and hence less strong heating rates during magnetic reconnection.

\begin{figure*}[ht!]
\centering
\includegraphics[width=1\textwidth]{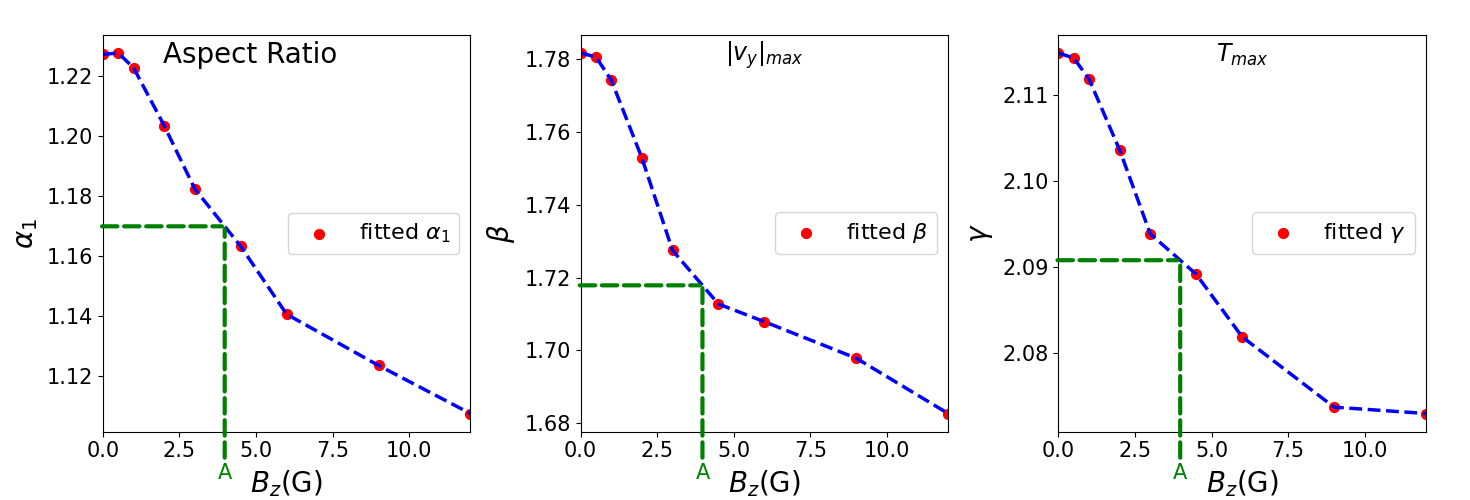}
\caption{Dependence of exponents $\alpha_1, \beta$, and $\gamma$, for the aspect ratio, absolute maximum upflow speed, and maximum temperature of the CS respectively, on the value of the coronal guide-field $B_z$. The red dots represent values of exponents obtained via fitting the temporal dependence of such quantities as shown in previous figures. The blue dashed lines represent the linear interpolation of $\alpha_1, \beta, \gamma$ for values of $B_{z}$. For all three quantities, the exponent decreases with increasing $B_{z}$. In the leftmost panel, the green dot represents the value of $\alpha$ estimated by \citet{ding2024disklimb} for an observed apparent CS on the solar disk originally reported by \citet{xue2020solar}, which is then subjected to a $B_{z}$ of 3.97~G. Green dots in the remaining panels are the backward-predicted values of $\beta$ and $\gamma$ for maximum upflow velocity and maximum temperature respectively using the value of $B_{z}$ denoted by $A$.}
\label{fig:alpha_vs_Bz_aspect_ratio}
\end{figure*}

\subsection{Relation between power-law exponents and guide-field for various physical quantities and comparison with observations}

Figure~\ref{fig:alpha_vs_Bz_aspect_ratio} shows the trend of variation of $\alpha_1,\beta,\gamma$ with the guide-field $B_z$ for the corresponding physical quantities: aspect ratio of the CS, maximum upflow velocity $v_{y, max}$, and maximum temperature $T_{max}$. For each quantity, $\alpha_1, \beta,\gamma$ are the best-fit power-law exponents of temporal variation of that specific quantity (see Figures~\ref{fig:aspect_ratio_curve_fit}-\ref{fig:max_vy}--\ref{fig:max_temp}). For the aspect ratio, we only consider the first phase $\alpha_1$, i.e., the rapid elongation and thinning phase of the CS, for comparison with the observed value of 1.17 \citep[][]{ding2024disklimb, xue2020solar}. This value does not lie in our computed range for $\alpha_2$. Such estimated values of the exponent are displayed as red dots in all four panels of Figure~\ref{fig:alpha_vs_Bz_aspect_ratio}. Blue dashed curves indicate the interpolated values of $\alpha$ for $B_{z}$ between 0 and 12~G at an equal spacing of 0.5~G, which suggest that the corresponding exponent decreases with increasing values of guide-field for these three measured quantities.

\begin{figure*}[ht!]
\centering
\includegraphics[width=\linewidth]{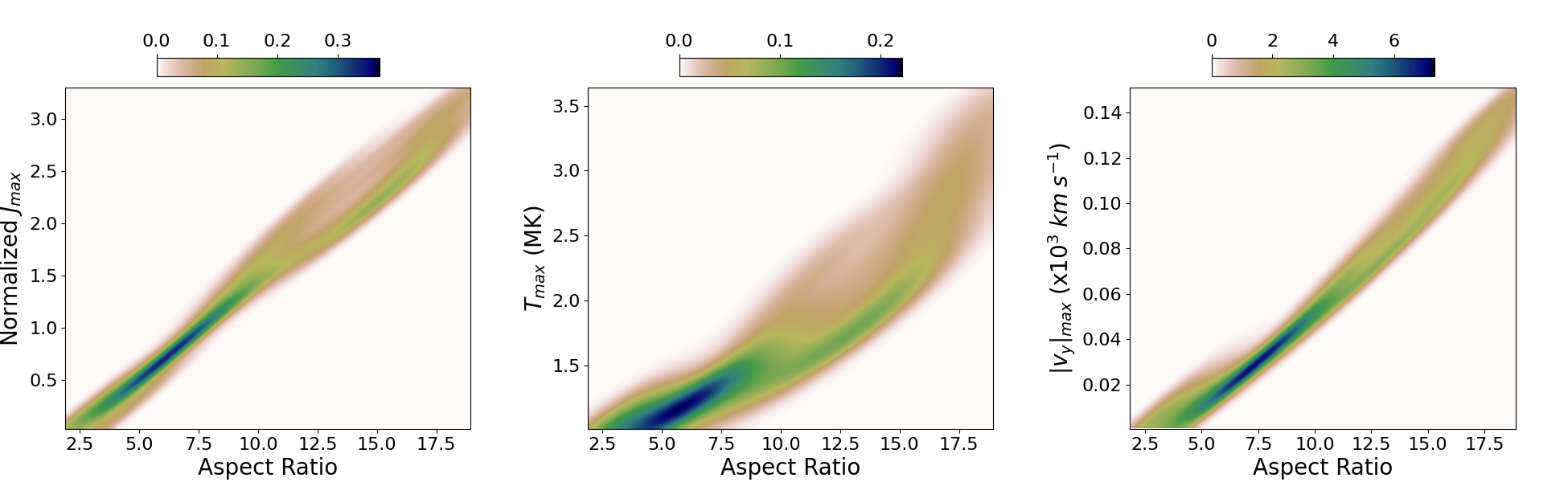}
\caption{Kernel Density Estimation as a visualization of the correlation of normalized $J_{max}$, $T_{max}$, and $|v_{y}|_{max}$ with the aspect ratio of the CS irrespective of the value of the guide-field and of the simulation time.} 
\label{fig:kde}

\end{figure*}

A direct comparison between theoretical or numerical current sheets and observed current sheets is not straightforward. In imaging observations, bright ray-like plasma structures are frequently identified as candidates for current sheets, but it is ambiguous whether these features are actually associated with the physical current sheets responsible for the magnetic-reconnection process under study. As discussed by \citet{ding2024disklimb}, even when an observed feature is physically associated with magnetic reconnection it is difficult to determine whether the structure fully represents the reconnecting current sheet or merely its surrounding emission pattern. Similarly, \citet{kwon2016} argued that post-CME white-light rays are associated with reconnection in the late stage of CMEs and do not necessarily represent the main flare-phase current sheet itself. These observational uncertainties necessitate a careful approach when comparing numerical results with remote-sensing data.

Following the prescription proposed by \citet{mondal2024CS}, \citet{ding2024disklimb} calculated various parameters that characterize the evolution of multiple on-disk as well as off-limb apparent CSs observed by SDO/AIA and SOHO/LASCO C2, such as Length ($L$), Width ($W$), aspect ratio ($L/W$), exponent ($\alpha$) with which aspect ratio changes with time, etc. Among these, the on-disk cases are more relevant to our present numerical model. \citet{ding2024disklimb} showed that these cases form within an X-type configuration and are associated with externally driven reconnection. This aligns with our setup, where the current sheet forms dynamically through the collapse of an X-type null under externally driven converging magnetic-field lines and the consequent response of the corona. They also found that on-disk current sheets evolve from bright points to elongated line-like structures, thinning in the late stage and exhibiting plasmoid-chain-like fine structures. Their samples show that on-disk CSs have $\alpha$ within 1.03 to 1.64 with aspect ratios being mostly within approximately 2--13. These properties are qualitatively consistent with our simulations, where the structure starts as a localized concentration, elongates, and undergoes thinning as the aspect ratio increases. Thus, we suggest that these on-disk candidates are plausible observational counterparts to the externally driven current sheets in our study.

The simulation results presented in this work demonstrate signatures of Sweet--Parker reconnection, as expected from resistive MHD, and 
are compared exclusively with on-disk apparent current sheets 
\citep[e.g.,][]{yang2014magnetic,li2016magnetic,xue2016observing,zhou2017study,yang2018observational,xue2020solar,hou2021formation,ding2022magnetic,yan2022fast,mondal2023reconnection}, which are formed within 
X-type magnetic configurations --- topologically consistent with the 
null-point setup employed here --- and are distinct from current 
sheets associated with eruptive flare--CME events.  
Lack of refinement in our simulations prohibits the generation of smaller-scale structures such as plasmoids, which are theoretically expected to occur at much higher Lundquist numbers $\approx~10^5$. They also observe the formation of plasmoid chain-like structures which cannot be reproduced in the current setup. The simulation setup also inhibits the onset of fast-reconnection processes, which are more likely in the corona.

We find that $\alpha$ estimated by \citet{ding2024disklimb} for the on-disk CS originally reported by \citet{xue2020solar} is around 1.17, and therefore it is within the range of $\alpha_1$ values estimated by us (see horizontal green dashed line in left panel of Figure~\ref{fig:alpha_vs_Bz_aspect_ratio}). Hence, assuming that the observed CS region has a similar resistivity and other background physical conditions as fixed in our simulation, we find that such a value of $\alpha$ corresponds to a guide-field $\approx$ 3.97~G (denoted via $A$ in all four panels of Figure~\ref{fig:alpha_vs_Bz_aspect_ratio}. Then, once we estimate the probable value of guide-field, we calculate at which rate, i.e., with which $\beta$ and $\gamma$, the upflow velocity and maximum temperature will increase with time under the aforementioned conditions (see green dots in top-right and two bottom panels of Figure~\ref{fig:alpha_vs_Bz_aspect_ratio}). In this way, we find predicted values of 1.72 and 2.09 for $\beta$ and $\gamma$ respectively.

\subsection{Potential correlations between various physical quantities independent of guide-field: A possibility of observational diagnosis}
So far, we have performed all the analyses knowing the value of $B_{z}$, since it is user-defined in our simulation. With a set of different observations of CS dynamics, we can find the evolution of the aspect ratio, maximum temperature from DEM analysis, and maximum upflow velocity for each event across time; but we have no reliable way to estimate $B_z$. In this section, we apply a statistical comparison between estimated observable quantities assuming no prior knowledge of the range of the underlying $B_{z}$, to understand whether any specific distributions are followed.

To this end, we estimate the kernel density distribution of maximum current density, maximum temperature, and maximum upflow velocity with the aspect ratio of the CS, see Figure~\ref{fig:kde}. We find that with an increase in the aspect ratio of the CS, current density, temperature, and upflow speed all increase irrespective of the value of the guide-field. This independence of the monotonic trends on the guide-field value suggests that such clustering, if observed, would primarily encode information about background plasma conditions (density, resistivity regime) rather than the guide field alone. Moreover, the density distribution of data-points is not uniformly distributed across all values of aspect ratios, among different quantities. Rather, they have higher densities at specific ranges of aspect ratio values. Such localization of data-point density may be a function of background and magnetic field conditions, resistivity etc. Therefore, producing such a distribution from a large set of observations, or observational proxies of aforementioned indicators may act as a probabilistic way to probe background coronal conditions. However, its practical realization requires a substantial exploration of parameter spaces within resistivity and magnetic field conditions, which is out-of-scope for the current studies.

\section{Conclusion} \label{sec:highlight}
We model the evolution of localized coronal CSs and ongoing magnetic reconnection within them, subjected to converging motion of opposite magnetic polarities \citep[][]{syntelis2019cancellation} at the base of the model corona, in the presence of weak and strong guide-fields, with respect to the in-plane reconnecting field. The reconnection driving is inspired by \citet{priest2018cancellation} and \citet{syntelis2019cancellation}, which for sufficiently strong magnetic-flux sources and underlying field strengths have proven to heat the QS and AR chromosphere sufficiently through fast reconnection \citep[][]{priest2018cancellation}. The model employed in this work utilizes a similar driver with weaker magnetic flux sources to study the evolution of localized CSs in purely coronal settings. A maximum temperature of two to three times the initial ambient temperature achieved within the CS during the later stages of reconnection show the capacity of localized CSs to locally heat the corona, even under the conservative 
conditions of uniform resistivity. 

We observe that the temporal evolution of the reconnection rate and aspect ratio can be divided into two phases, with different exponents of power-law fits in the two time intervals. The first phase, in which there occurs a rapid growth in both the reconnection rate and aspect ratio, demonstrates a downward trend in the fitted exponent with increasing guide-field. However, the second phase demonstrates an increasing trend in the power-law fitted exponent. This implies that an increasing guide-field hinders the initial rapid growth phase of aspect ratio and reconnection rate, but acts to enhance the rate of reconnection in the second phase. The evolution of observables such as the maximum temperature and bidirectional outflow also displays a hindered growth for increasing guide-field.

The decreasing trend of $\delta_1$ with increasing guide-field strength found in our study are consistent with previous studies of guide-field effects on antiparallel reconnection using neutral Harris-type current sheets \citep[e.g.,][]{pritchett2004three, huba2005hall, huang2016turbulent}. They reported that the guide-field has a decremental effect on the rate of reconnection and plasma energization. An important distinction in our simulations from previous studies is the self-consistent formation of CSs with guide-fields that are present from the start. Here, we observe a trend similar to that of a conventional externally driven reconnection problem, even when the resistivity is independent of time and space.

In contrast to the early-stage behavior, $\delta_2$ shows an increasing trend with increasing guide-field strength in the later phase, although the onset of this phase, $\theta_2$, is delayed to later times for stronger guide fields. 
The time-windows of $\alpha_2$ and $\delta_2$ correspond to drops in the average guide-field strength within the CS region, the drop-rate being inversely proportional to $B_z$ (see Appendix~\ref{sec:app2} and Figure~\ref{fig:appendix_bz}) and rising growth-rates of inflow velocities into the CS, with growth-rates directly proportional to $B_z$ in later phases of reconnection (see Appendix~\ref{sec:app2} and Figure~\ref{fig:appendix_vx}). This indicates that inflows into the CS essentially overcome any hindrance posed by $B_z$, leading to the later phase characterized by faster temporal growth. This delayed commencement and evolution of the second phase at stronger guide-field strengths could potentially be interpreted as allowing magnetic energy to accumulate for a longer period before the onset of the later phase, consistent with the general picture of magnetic-energy storage prior to its release through reconnection \citep[e.g.,][]{pontin2022magnetic}.

Overall, the calculated values of length, width, and aspect ratio of the current sheets correspond well to the observed ranges of values obtained by \citet{ding2024disklimb} for on-disk apparent current sheets. Using the value of $\alpha$ calculated by \citet{ding2024disklimb}, from the observations of \citet{xue2020solar}, we interpolate to find the value of corresponding guide-field from Figure~\ref{fig:alpha_vs_Bz_aspect_ratio}, and then we backtrack to calculate the exponents of temporal variation for maximum upflow velocity and maximum temperature within the current sheet. This is a test of the observed on-disk coronal current sheet going into the thinning and reconnection, which now falls in the realm of our theoretical model calculations. However, more such observed CS and their analyses are required to fit in to the parametric space of the present numerical model. More parametric studies using our numerical model can be done in the future assimilating such a wide cohort of the observational image data of CS in place.

We also analyze the statistical correlation between aspect ratio and reconnection rate proxies such as $\eta J_{max}$. temperature, plasma upflow speed . This analysis demonstrates a high correlation between reconnection indicators and aspect ratio without any knowledge on value of guide-fields associated with any data point. An important implication of 2.5D MHD simulations is to effectively approach the limit of zero magnetic shear in the ignorable direction ($z$, in our case). It has been shown that CSs with (2.5D) and without (2D) a guide-field cannot be in equilibrium while possessing finite thickness and length \citep[][]{klimchucketal2023}. Simulations in three dimensions are essential to understand the role of magnetic tension produced by curved guide fields, which will be tackled later on. In future studies, we aim to understand the effects of anomalous resistivity and spatially nonuniform guide-fields on the temporal variation of the parameters studied here, and how the correlations between reconnection indicators and aspect ratios are affected.
The impact of complex 3D magnetic fields on CS energetics and dynamics also remains as a topic for future studies.

\begin{acknowledgements}
We acknowledge the National Supercomputing Mission (NSM) for providing computing resources of 'PARAM Shivay' at Indian Institute of Technology (BHU), Varanasi, which is implemented by C-DAC and supported by the Ministry of Electronics and Information Technology (MeitY) and Department of Science and Technology (DST), Government of India.
We also thank the open-source PLUTO 4.4 code for providing a user-friendly framework. 
S.M.\ acknowledges Prime Minister Research Fellowship of India for providing financial support. A.K.S.\  acknowledges ISRO grant (DS/2B-13012(2)/26/2022-Sec.2) for supporting his scientific research. D.Y.\ was supported by the National Natural Science Foundation of China (NSFC: 12473050), the Guangdong Natural Science Funds for Distinguished Young Scholars (2023B1515020049), the Shenzhen Science and Technology Project (JCYJ20240813104805008), and the Specialized Research Fund for State Key Laboratory of Solar Activity and Space Weather. F.B.\ acknowledges support from the FED-tWIN programme (profile Prf-2020-004, project ``ENERGY''), issued by BELSPO, and from the FWO Junior Research Project G020224N granted by the Research Foundation -- Flanders (FWO). X.P.Z. is supported by the Natural Science Foundation of China (Grant No. 12303062) and the Sichuan Science and Technology Program (Grant No. 2026NSFSC0020). R.Y.K. acknowledges support by basic research funding from the Korea Astronomy and Space Science Institute (KASI; KASI2026183005).
\end{acknowledgements}

\bibliographystyle{aa}  
\bibliography{aa60261-26}

\begin{thebibliography}{61}
\expandafter\ifx\csname natexlab\endcsname\relax\def\natexlab#1{#1}\fi

\bibitem[{{Alissandrakis} \& {Gary}(2021)}]{2Alissandrakis2021cormag}
{Alissandrakis}, C.~E. \& {Gary}, D.~E. 2021, Frontiers in Astronomy and Space Sciences, 7, 77

\bibitem[{Balsara \& Spicer(1999)}]{balsara1999staggered}
Balsara, D.~S. \& Spicer, D.~S. 1999, Journal of Computational Physics, 149, 270

\bibitem[{Beg {et~al.}(2022)Beg, Russell, \& Hornig}]{beg2022}
Beg, R., Russell, A. J.~B., \& Hornig, G. 2022, The Astrophysical Journal, 940, 94

\bibitem[{{Benz}(2017)}]{benz2017flare}
{Benz}, A.~O. 2017, Living Reviews in Solar Physics, 14, 2

\bibitem[{{Birn} {et~al.}(2005){Birn}, {Galsgaard}, {Hesse}, {Hoshino}, {Huba}, {Lapenta}, {Pritchett}, {Schindler}, {Yin}, {B{\"u}chner}, {Neukirch}, \& {Priest}}]{birn2005forced}
{Birn}, J., {Galsgaard}, K., {Hesse}, M., {et~al.} 2005, \grl, 32, L06105

\bibitem[{{Birn} \& {Hesse}(2010)}]{birn2010energy}
{Birn}, J. \& {Hesse}, M. 2010, Physics of Plasmas, 17, 012109

\bibitem[{{Brooks} {et~al.}(2021){Brooks}, {Warren}, \& {Landi}}]{brooks2021cormag}
{Brooks}, D.~H., {Warren}, H.~P., \& {Landi}, E. 2021, \apjl, 915, L24

\bibitem[{Chen {et~al.}(2020)Chen, Shen, Gary, Reeves, Fleishman, Yu, Guo, Krucker, Lin, Nita, {et~al.}}]{chen2020measurement}
Chen, B., Shen, C., Gary, D.~E., {et~al.} 2020, Nature Astronomy, 4, 1140

\bibitem[{{Cheng} {et~al.}(2020){Cheng}, {Zhang}, {Kliem}, {T{\"o}r{\"o}k}, {Xing}, {Zhou}, {Inhester}, \& {Ding}}]{cheng2020initiation}
{Cheng}, X., {Zhang}, J., {Kliem}, B., {et~al.} 2020, \apj, 894, 85

\bibitem[{Dahlin {et~al.}(2022)Dahlin, Antiochos, Qiu, \& DeVore}]{dahlin2022variability}
Dahlin, J.~T., Antiochos, S.~K., Qiu, J., \& DeVore, C.~R. 2022, The Astrophysical Journal, 932, 94

\bibitem[{Daughton {et~al.}(2011)Daughton, Roytershteyn, Karimabadi, Yin, Albright, Bergen, \& Bowers}]{daughton2011role}
Daughton, W., Roytershteyn, V., Karimabadi, H., {et~al.} 2011, Nature Physics, 7, 539

\bibitem[{{Ding} \& {Zhang}(2024)}]{ding2024disklimb}
{Ding}, T. \& {Zhang}, J. 2024, \apj, 974, 104

\bibitem[{{Ding} {et~al.}(2022){Ding}, {Zhang}, \& {Hong}}]{ding2022magnetic}
{Ding}, T., {Zhang}, J., \& {Hong}, J. 2022, \apjl, 933, L38

\bibitem[{Evans \& Hawley(1988)}]{evans1988simulation}
Evans, C.~R. \& Hawley, J.~F. 1988, Simulation of Magnetohydrodynamic Flows: A Constrained Transport Model., 332: 659

\bibitem[{F{\ae}rder {et~al.}(2023)F{\ae}rder, N{\'o}brega-Siverio, \& Carlsson}]{faerder2023comparative}
F{\ae}rder, {\O}.~H., N{\'o}brega-Siverio, D., \& Carlsson, M. 2023, arXiv preprint arXiv:2305.18895

\bibitem[{{Gopalswamy} {et~al.}(2024){Gopalswamy}, {Micha{\l}ek}, {Yashiro}, {M{\"a}kel{\"a}}, {Akiyama}, {Xie}, \& {Vourlidas}}]{gopalswamy2024soho}
{Gopalswamy}, N., {Micha{\l}ek}, G., {Yashiro}, S., {et~al.} 2024, arXiv e-prints, arXiv:2407.04165

\bibitem[{Harten {et~al.}(1983)Harten, Lax, \& Leer}]{harten1983upstream}
Harten, A., Lax, P.~D., \& Leer, B.~v. 1983, SIAM review, 25, 35

\bibitem[{{Hou} {et~al.}(2021){Hou}, {Tian}, {Chen}, {Zhu}, {Huang}, {Bai}, {He}, {Song}, \& {Xia}}]{hou2021formation}
{Hou}, Z., {Tian}, H., {Chen}, H., {et~al.} 2021, \apj, 915, 39

\bibitem[{{Huang} \& {Bhattacharjee}(2016)}]{huang2016turbulent}
{Huang}, Y.-M. \& {Bhattacharjee}, A. 2016, \apj, 818, 20

\bibitem[{{Huba}(2005)}]{huba2005hall}
{Huba}, J.~D. 2005, Physics of Plasmas, 12, 012322

\bibitem[{{Hudson}(2011)}]{hudson2011global}
{Hudson}, H.~S. 2011, \ssr, 158, 5

\bibitem[{Izquierdo-Guzm{\'a}n {et~al.}(2023)Izquierdo-Guzm{\'a}n, Gonz{\'a}lez-Avil{\'e}s, \& Guzm{\'a}n}]{izquierdo2023}
Izquierdo-Guzm{\'a}n, I., Gonz{\'a}lez-Avil{\'e}s, J.~J., \& Guzm{\'a}n, F.~S. 2023, Solar Physics, 298, 109

\bibitem[{Janhunen(2000)}]{janhunen2000positive}
Janhunen, P. 2000, Journal of Computational Physics, 160, 649

\bibitem[{Klimchuk {et~al.}(2023)}]{klimchucketal2023}
Klimchuk, J.~A. {et~al.} 2023, Frontiers in Physics, Volume 11 - 2023

\bibitem[{Kwon {et~al.}(2016)Kwon, Vourlidas, \& Webb}]{kwon2016}
Kwon, R.-Y., Vourlidas, A., \& Webb, D. 2016, The Astrophysical Journal, 826, 94

\bibitem[{{Li} {et~al.}(2016){Li}, {Zhang}, {Peter}, {Priest}, {Chen}, {Guo}, {Chen}, \& {Mackay}}]{li2016magnetic}
{Li}, L., {Zhang}, J., {Peter}, H., {et~al.} 2016, Nature Physics, 12, 847

\bibitem[{{Mignone} {et~al.}(2007){Mignone}, {Bodo}, {Massaglia}, {Matsakos}, {Tesileanu}, {Zanni}, \& {Ferrari}}]{mignone2007pluto}
{Mignone}, A., {Bodo}, G., {Massaglia}, S., {et~al.} 2007, \apjs, 170, 228

\bibitem[{{Mondal} {et~al.}(2025{\natexlab{a}}){Mondal}, {Bairagi}, \& {Srivastava}}]{mondal2025CS2}
{Mondal}, S., {Bairagi}, A., \& {Srivastava}, A.~K. 2025{\natexlab{a}}, \apj, 979, 207

\bibitem[{{Mondal} {et~al.}(2023){Mondal}, {Srivastava}, {Mishra}, {Sangal}, {Kayshap}, {Guo}, {Pontin}, {Uritsky}, {Ofman}, {Wang}, \& {Yuan}}]{mondal2023reconnection}
{Mondal}, S., {Srivastava}, A.~K., {Mishra}, S.~K., {et~al.} 2023, \apj, 953, 84

\bibitem[{{Mondal} {et~al.}(2025{\natexlab{b}}){Mondal}, {Srivastava}, {Pontin}, \& {Priest}}]{2025ApJ...989..222M}
{Mondal}, S., {Srivastava}, A.~K., {Pontin}, D.~I., \& {Priest}, E.~R. 2025{\natexlab{b}}, \apj, 989, 222

\bibitem[{{Mondal} {et~al.}(2024){Mondal}, {Srivastava}, {Pontin}, {Yuan}, \& {Priest}}]{mondal2024CS}
{Mondal}, S., {Srivastava}, A.~K., {Pontin}, D.~I., {Yuan}, D., \& {Priest}, E.~R. 2024, \apj, 963, 139

\bibitem[{{Moore} {et~al.}(2001){Moore}, {Sterling}, {Hudson}, \& {Lemen}}]{moore2001onset}
{Moore}, R.~L., {Sterling}, A.~C., {Hudson}, H.~S., \& {Lemen}, J.~R. 2001, \apj, 552, 833

\bibitem[{{Parker}(1957)}]{parker1957sweet}
{Parker}, E.~N. 1957, \jgr, 62, 509

\bibitem[{{Pontin} \& {Priest}(2022)}]{pontin2022magnetic}
{Pontin}, D.~I. \& {Priest}, E.~R. 2022, Living Reviews in Solar Physics, 19, 1

\bibitem[{Pontin {et~al.}(2023)Pontin, Priest, Chitta, \& Titov}]{Pontin_2024}
Pontin, D.~I., Priest, E.~R., Chitta, L.~P., \& Titov, V.~S. 2023, The Astrophysical Journal, 960, 51

\bibitem[{{Priest}(2014)}]{priest2014magnetohydrodynamics}
{Priest}, E. 2014, {Magnetohydrodynamics of the Sun}

\bibitem[{{Priest} {et~al.}(2018){Priest}, {Chitta}, \& {Syntelis}}]{priest2018cancellation}
{Priest}, E.~R., {Chitta}, L.~P., \& {Syntelis}, P. 2018, \apjl, 862, L24

\bibitem[{{Pritchett} \& {Coroniti}(2004)}]{pritchett2004three}
{Pritchett}, P.~L. \& {Coroniti}, F.~V. 2004, Journal of Geophysical Research (Space Physics), 109, A01220

\bibitem[{{Qiu} {et~al.}(2010){Qiu}, {Liu}, {Hill}, \& {Kazachenko}}]{qiu2010reconnection}
{Qiu}, J., {Liu}, W., {Hill}, N., \& {Kazachenko}, M. 2010, \apj, 725, 319

\bibitem[{Robinson {et~al.}(2022)Robinson, Carlsson, \& Aulanier}]{robinson2022incoherent}
Robinson, R.~A., Carlsson, M., \& Aulanier, G. 2022, Astronomy \& Astrophysics, 668, A177

\bibitem[{Schreier {et~al.}(2010)Schreier, Swisdak, Drake, \& Cassak}]{schreier2010three}
Schreier, R., Swisdak, M., Drake, J., \& Cassak, P. 2010, Physics of Plasmas, 17

\bibitem[{Sen \& Keppens(2022)}]{Sen2022}
Sen, S. \& Keppens, R. 2022, Astronomy \& Astrophysics, 666, A28

\bibitem[{{Shay} {et~al.}(2014){Shay}, {Haggerty}, {Phan}, {Drake}, {Cassak}, {Wu}, {Oieroset}, {Swisdak}, \& {Malakit}}]{shay2014electron}
{Shay}, M.~A., {Haggerty}, C.~C., {Phan}, T.~D., {et~al.} 2014, Physics of Plasmas, 21, 122902

\bibitem[{{Smitha} {et~al.}(2017){Smitha}, {Anusha}, {Solanki}, \& {Riethm{\"u}ller}}]{smitha2017estimation}
{Smitha}, H.~N., {Anusha}, L.~S., {Solanki}, S.~K., \& {Riethm{\"u}ller}, T.~L. 2017, \apjs, 229, 17

\bibitem[{Snow {et~al.}(2018)Snow, Botha, McLaughlin, \& Hillier}]{snow2018}
Snow, B., Botha, G. J.~J., McLaughlin, J.~A., \& Hillier, A. 2018, Astronomy \& Astrophysics, 609, A100

\bibitem[{{Solanki} {et~al.}(2010){Solanki}, {Barthol}, {Danilovic}, {Feller}, {Gandorfer}, {Hirzberger}, {Riethm{\"u}ller}, {Sch{\"u}ssler}, {Bonet}, {Mart{\'\i}nez Pillet}, {del Toro Iniesta}, {Domingo}, {Palacios}, {Kn{\"o}lker}, {Bello Gonz{\'a}lez}, {Berkefeld}, {Franz}, {Schmidt}, \& {Title}}]{solanki2010sunrise}
{Solanki}, S.~K., {Barthol}, P., {Danilovic}, S., {et~al.} 2010, \apjl, 723, L127

\bibitem[{{Solanki} {et~al.}(2017){Solanki}, {Riethm{\"u}ller}, {Barthol}, {Danilovic}, {Deutsch}, {Doerr}, {Feller}, {Gandorfer}, {Germerott}, {Gizon}, {Grauf}, {Heerlein}, {Hirzberger}, {Kolleck}, {Lagg}, {Meller}, {Tomasch}, {van Noort}, {Blanco Rodr{\'\i}guez}, {Gasent Blesa}, {Balaguer Jim{\'e}nez}, {Del Toro Iniesta}, {L{\'o}pez Jim{\'e}nez}, {Orozco Suarez}, {Berkefeld}, {Halbgewachs}, {Schmidt}, {{\'A}lvarez-Herrero}, {Sabau-Graziati}, {P{\'e}rez Grande}, {Mart{\'\i}nez Pillet}, {Card}, {Centeno}, {Kn{\"o}lker}, \& {Lecinski}}]{solanki2017second}
{Solanki}, S.~K., {Riethm{\"u}ller}, T.~L., {Barthol}, P., {et~al.} 2017, \apjs, 229, 2

\bibitem[{{Srivastava} {et~al.}(2025){Srivastava}, {Mondal}, {Priest}, {Mishra}, {Pontin}, {Kwon}, {Yuan}, {Murawski}, \& {Asai}}]{2025ApJ...984...36S}
{Srivastava}, A.~K., {Mondal}, S., {Priest}, E.~R., {et~al.} 2025, \apj, 984, 36

\bibitem[{{Syntelis} {et~al.}(2019){Syntelis}, {Priest}, \& {Chitta}}]{syntelis2019cancellation}
{Syntelis}, P., {Priest}, E.~R., \& {Chitta}, L.~P. 2019, \apj, 872, 32

\bibitem[{Tharp {et~al.}(2012)Tharp, Yamada, Ji, Lawrence, Dorfman, Myers, \& Yoo}]{Tharp2012}
Tharp, T.~D., Yamada, M., Ji, H., {et~al.} 2012, Phys. Rev. Lett., 109, 165002

\bibitem[{{Trattner} {et~al.}(2007){Trattner}, {Mulcock}, {Petrinec}, \& {Fuselier}}]{trattner2007location}
{Trattner}, K.~J., {Mulcock}, J.~S., {Petrinec}, S.~M., \& {Fuselier}, S.~A. 2007, \grl, 34, L03108

\bibitem[{{Webb} \& {Howard}(2012)}]{webb2012coronal}
{Webb}, D.~F. \& {Howard}, T.~A. 2012, Living Reviews in Solar Physics, 9, 3

\bibitem[{{Xue} {et~al.}(2016){Xue}, {Yan}, {Cheng}, {Yang}, {Su}, {Kliem}, {Zhang}, {Liu}, {Bi}, {Xiang}, {Yang}, \& {Zhao}}]{xue2016observing}
{Xue}, Z., {Yan}, X., {Cheng}, X., {et~al.} 2016, Nature Communications, 7, 11837

\bibitem[{{Xue} {et~al.}(2020){Xue}, {Yan}, {Yang}, {Wang}, {Li}, \& {Zhao}}]{xue2020solar}
{Xue}, Z., {Yan}, X., {Yang}, L., {et~al.} 2020, \aap, 633, A121

\bibitem[{{Yamaleev} \& {Carpenter}(2009)}]{yamaleev2009weno}
{Yamaleev}, N.~K. \& {Carpenter}, M.~H. 2009, Journal of Computational Physics, 228, 4248

\bibitem[{{Yan} {et~al.}(2022){Yan}, {Xue}, {Jiang}, {Priest}, {Kliem}, {Yang}, {Wang}, {Kong}, {Song}, {Feng}, \& {Liu}}]{yan2022fast}
{Yan}, X., {Xue}, Z., {Jiang}, C., {et~al.} 2022, Nature Communications, 13, 640

\bibitem[{{Yang} {et~al.}(2018){Yang}, {Yang}, {Bi}, {Hong}, {Li}, {Xu}, \& {Chen}}]{yang2018observational}
{Yang}, B., {Yang}, J., {Bi}, Y., {et~al.} 2018, \apj, 861, 135

\bibitem[{{Yang} {et~al.}(2015){Yang}, {Zhang}, \& {Xiang}}]{yang2014magnetic}
{Yang}, S., {Zhang}, J., \& {Xiang}, Y. 2015, \apjl, 798, L11

\bibitem[{{Yokoyama} \& {Shibata}(2001)}]{yokoyama2001flare}
{Yokoyama}, T. \& {Shibata}, K. 2001, \apj, 549, 1160

\bibitem[{{Zhou} {et~al.}(2017){Zhou}, {Zhang}, {Wang}, \& {Wheatland}}]{zhou2017study}
{Zhou}, G.~P., {Zhang}, J., {Wang}, J.~X., \& {Wheatland}, M.~S. 2017, \apjl, 851, L1

\bibitem[{{Zweibel} \& {Yamada}(2016)}]{zweibel2016perspectives}
{Zweibel}, E.~G. \& {Yamada}, M. 2016, Proceedings of the Royal Society of London Series A, 472, 20160479

\end{thebibliography}

\begin{appendix}

\onecolumn
\section{Choice of data windows for power-law fitting}\label{sec:app1}

\begin{figure*}[ht!]
\centering
\includegraphics[width=\linewidth]{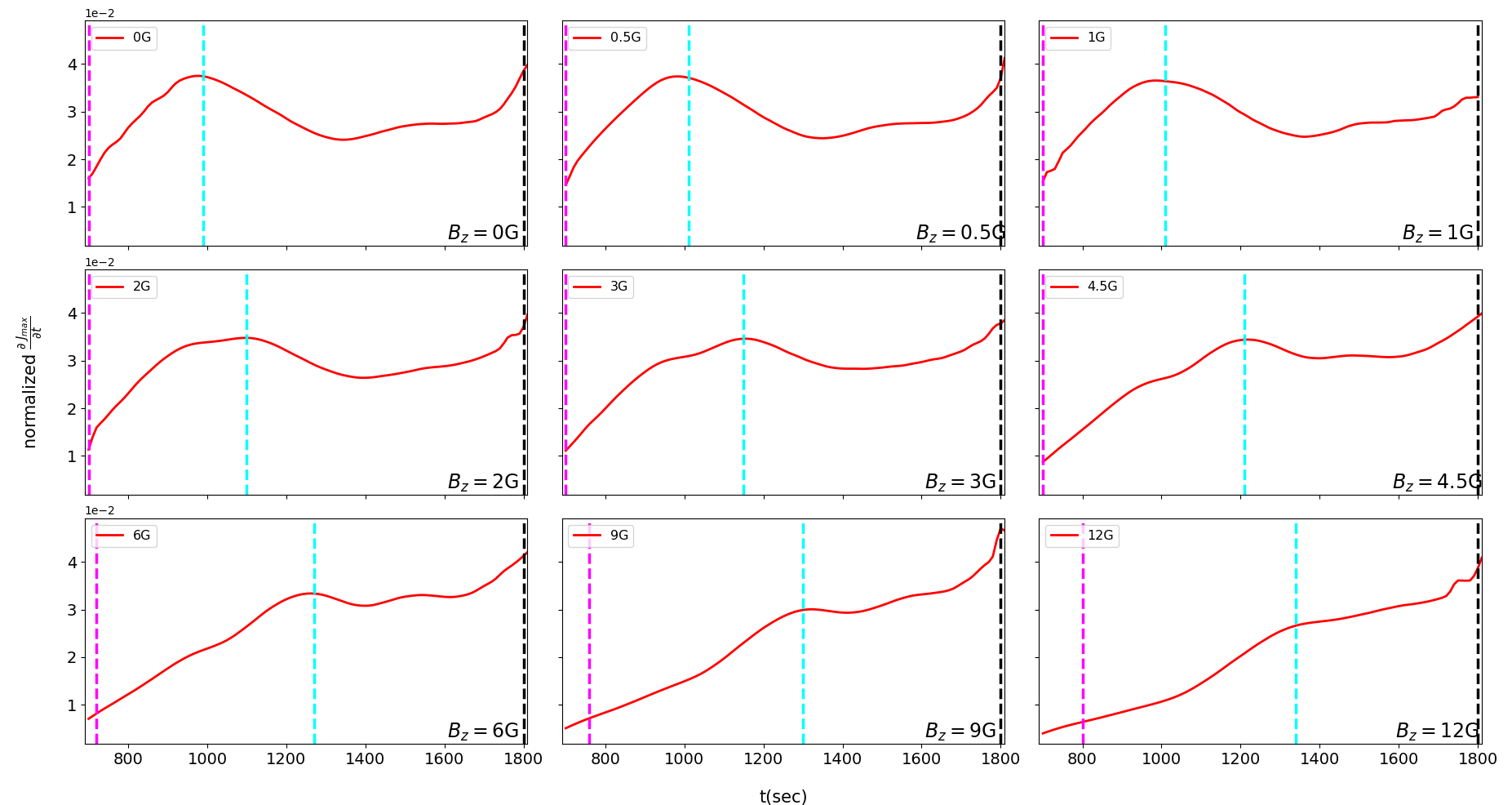}
\caption{Temporal evolution of the time derivative of $J_{max}$ (or reconnection rate) for different guide-field strengths $B_z$. $\delta_1$ is calculated for the time interval $\theta_1$--$\theta_2$~s, and $\delta_2$ is calculated for the time interval $\theta_2$--$\theta_3$~s. The vertical dashed magenta, cyan, and black lines denote $t =\theta_1$, $\theta_2$, and $\theta_3$~s respectively.
\label{fig:appendix_eta_j_max}
}
\end{figure*}

\begin{figure*}[h!]
\centering
\includegraphics[width=\linewidth]{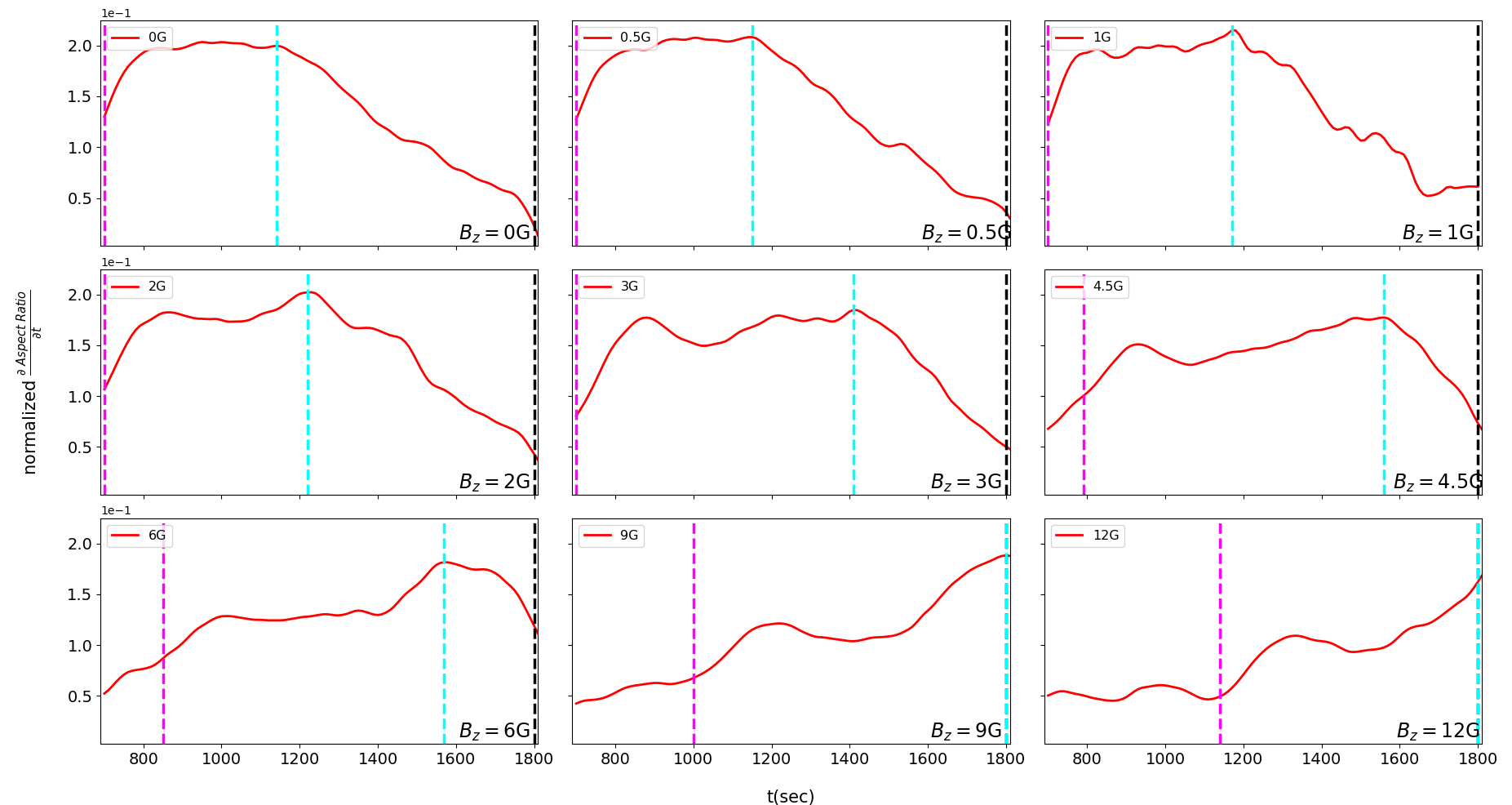}
\caption{Temporal evolution of the time derivative of the CS aspect ratio for increasing strength of the guide-field $B_z$. $\alpha_1$ is calculated for the time interval $\phi_1$--$\phi_2$~s , and $\alpha_2$ is calculated for the time interval $\phi_2$--$\phi_3$~s. The vertical dashed magenta, cyan, and black lines denote $t = \phi_1$, $\phi_2$, and $\phi_3$~s, respectively.
\label{fig:appendix_aspect_ratio}
}
\end{figure*}

\begin{figure*}[ht!]
\centering
\includegraphics[width=\linewidth]{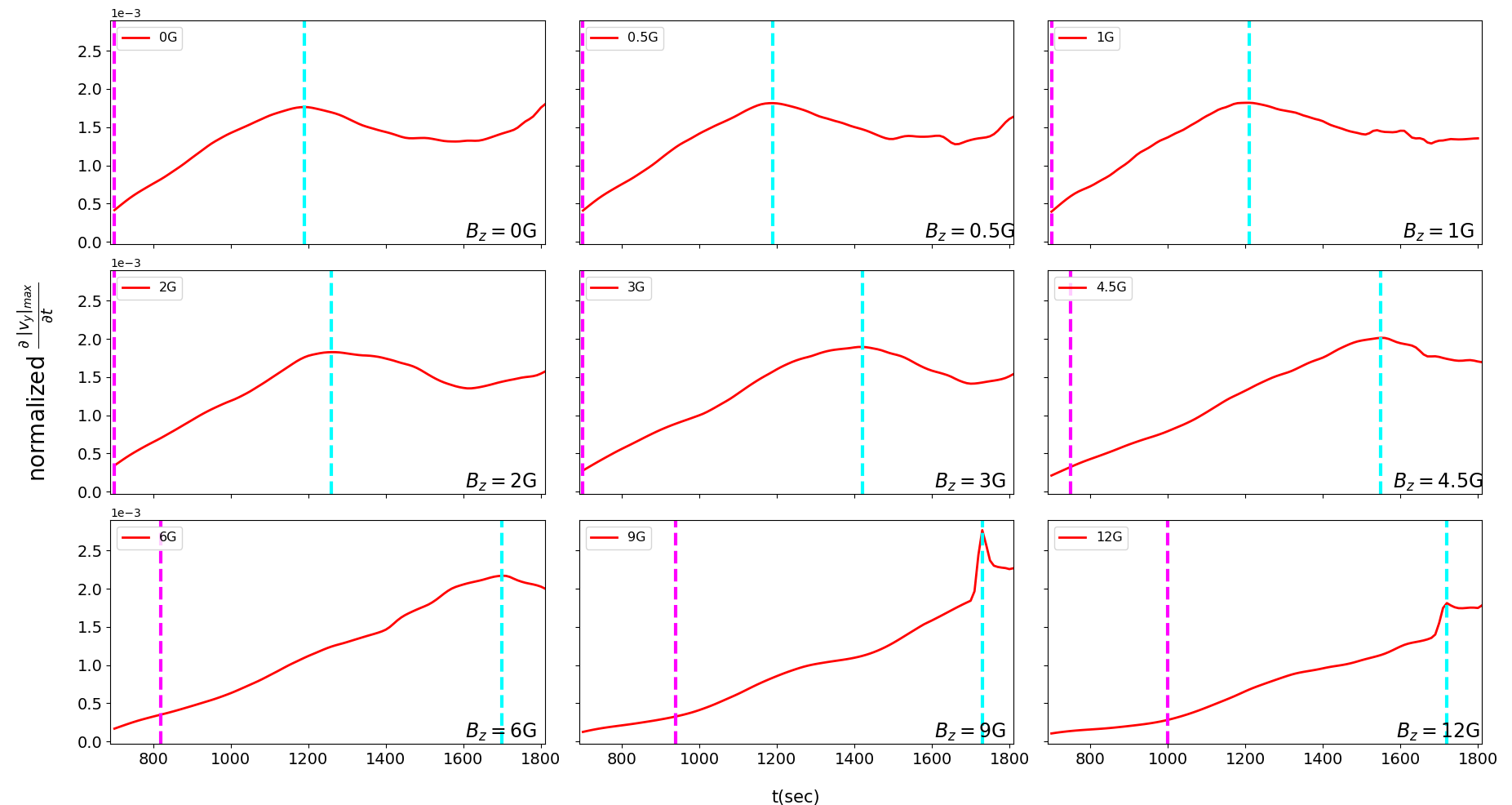}
\caption{Temporal evolution of the derivative of $|v_y|_{max}$. The fitting window is the time interval $\psi_1$--$\psi_2$~s, where the slope of the corresponding plots in Figure~\ref{fig:appendix_vy_max} increases monotonically. It can be seen that with increasing $B_z$, the time interval of the region where the derivative decreases is reduced. In stronger guide-fields, the dashed magenta line is shifted to later times, to account for an observable delay in the rapid increase of $|v_y|_{max}$, which is also apparent in $T_{max}$, aspect ratio, and $|v_y|_{max}$. The vertical dashed magenta and cyan lines denote $t = \psi_1$ and $\psi_2$~s respectively.
\label{fig:appendix_vy_max}
}
\end{figure*}

\begin{figure*}[ht!]
\centering
\includegraphics[width=\linewidth]{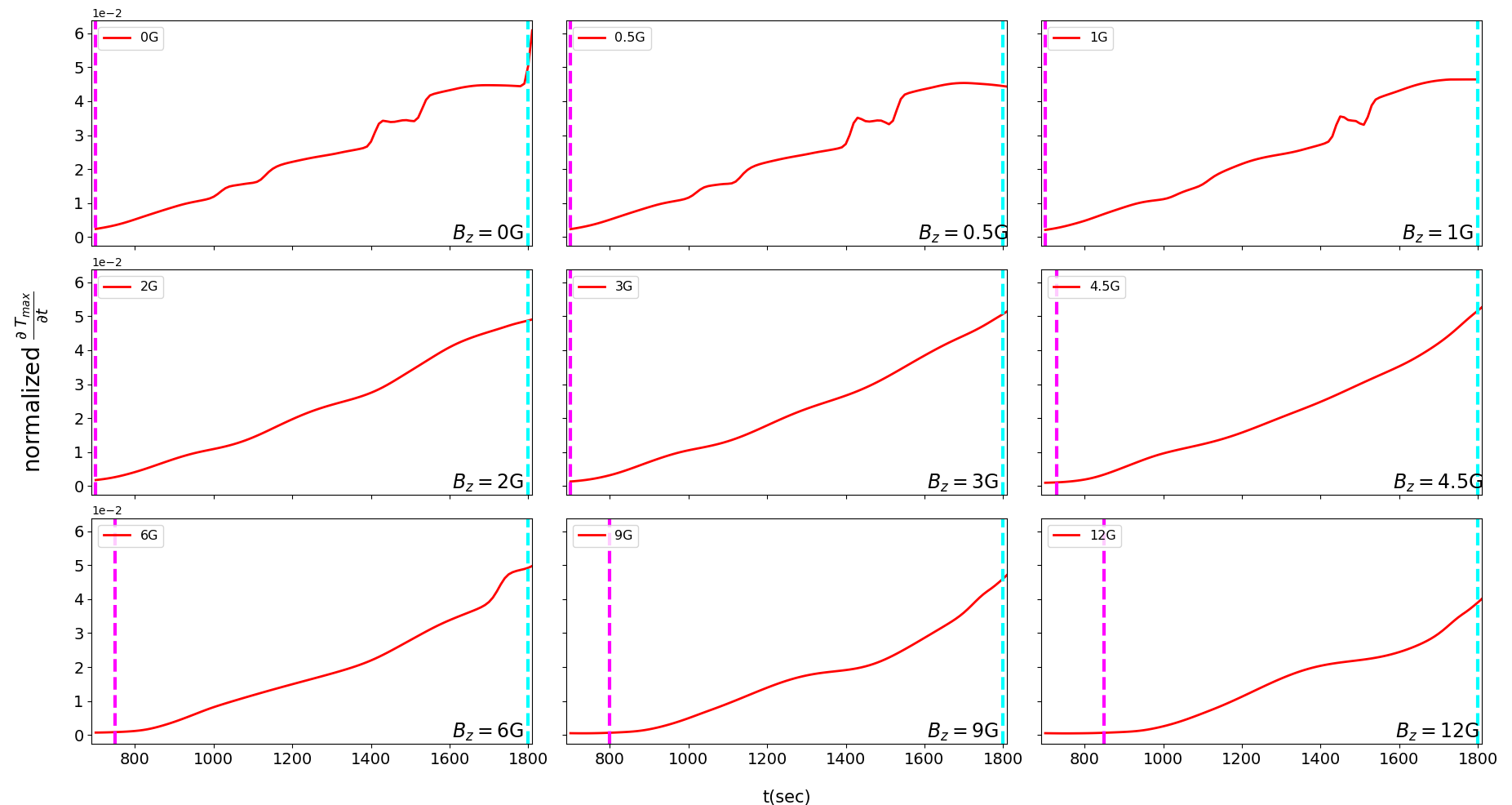}
\caption{Temporal evolution of the derivative of $T_{max}$. The time interval for fitting, $\chi_1$--$\chi_2$~s, is the interval where the slope of the corresponding plots in Figure~\ref{fig:max_temp} increases monotonically, which for most cases here covers the entire period from $t = 700$~s to $t = 1800$~s, with some exceptions for stronger guide-fields. High values of $B_z$ have a delaying effect on the rapid increase in $T_{max}$, which can increase errors in fitting. This is accounted for by shifting the magenta dashed line towards later times accordingly. The vertical dashed magenta and cyan lines denote $t =\chi_1$ and $\chi_2$~s respectively. 
\label{fig:appendix_Tmax}
}
\end{figure*}

As observed in Sections~\ref{sec:3.2}, \ref{sec:aspectratio}, \ref{sec:vymax}, and \ref{sec:tmax}, we have defined timestamps (in s) $\theta_1$, $\theta_2$, and $\theta_3$ in Figures~\ref{fig:etajmax} and \ref{fig:appendix_eta_j_max}; $\phi_1$, $\phi_2$, and $\phi_3$ in Figures~\ref{fig:aspect_ratio_curve_fit} and \ref{fig:appendix_aspect_ratio}; $\psi_1$ and $\psi_2$ in Figures~\ref{fig:max_vy} and \ref{fig:appendix_vy_max}; and $\chi_1$ and $\chi_2$ in Figures~\ref{fig:max_temp} and \ref{fig:appendix_Tmax},  for the reconnection rate, aspect ratio of the CS, maximum upflow velocity, and maximum temperature within the CS respectively. This is done in order to define the time intervals over which data is fitted according to a power-law (Equations~\eqref{eqn:delta}, \eqref{eqn:alpha}, \eqref{eqn:beta}, and \eqref{eqn:gamma} respectively), and exponents of best fit are measured. The reason for this approach is the inability of a single power-law fit to explain the behavior of $J_{max}$ and the aspect ratio of the CS in both phases, resulting in an inadequate fit. To rectify this, we differentiate all the quantities with respect to time, and apply a criterion for choosing the relevant timestamps based on intervals in which a quantity undergoes a rapid or slow change. Figures~\ref{fig:appendix_eta_j_max}, \ref{fig:appendix_aspect_ratio}, \ref{fig:appendix_vy_max}, and \ref{fig:appendix_Tmax} depict the variation of the time derivatives of $J_{max}$, Aspect Ratio, $|v_y|_{max}$, and $T_{max}$.

In Figure~\ref{fig:etajmax}, the evolution of the maximum current density within the CS (and hence the reconnection rate) can be divided into two distinct phases, each characterized by different values of the fitting exponent, $\delta_1$ and $\delta_2$. This division is based on the time when the derivative of $J_{max}$ starts decreasing or exhibits a slower change following a phase of steady and rapid increase, as indicated by the dashed cyan line ($t= \theta_2$~s) in Figure~\ref{fig:appendix_eta_j_max}. The dashed magenta line ($t = \theta_1$~s), marking the start of the fitting window for the first phase ($\delta_1$), is fixed at $t = 700$~s for most values of $B_z$. However, for higher guide-field strengths (6, 9, and 12~G), this line is shifted to later times to account for the observed delaying effect in the onset of rapid increase, which affects the fit quality. The dashed black line ($t=\theta_3$~s), representing the end of the fitting window for the second phase ($\delta_2$), is fixed at $t = 1800$~s across all cases. This results in the calculation of two exponents, $\delta_1$ and $\delta_2$, for the first and second phases, respectively. In Figure~\ref{fig:appendix_eta_j_max}, it can be observed that for $t>\theta_2$~s the region of slower change (decreasing derivative part of the latter time window) flattens over time with increasing $B_z$, where it becomes an inflection point around $B_z = 9$~G. This translates into an increasing trend for $\delta_2$, ranging from less than unity to greater than unity, as seen in Figure~\ref{fig:etajmax}. 

In Figure~\ref{fig:aspect_ratio_curve_fit}, in order to characterize the growth phase of the CS using the aspect ratio, we follow a similar strategy as prescribed for $J_{max}$ in Figure~\ref{fig:appendix_eta_j_max}. We define two time intervals $\phi_1$--$\phi_2$~s (between magenta and cyan dashed vertical lines) and $\phi_2$--$\phi_3$~s (between cyan and black vertical dashed lines) for calculating two exponents for the best power-law fit, $\alpha_1$ and $\alpha_2$ respectively. For the majority of $B_z$ values, the start of the fitting window $t = \phi_1$~s for $\alpha_1$ (dashed magenta line in Figure~\ref{fig:appendix_aspect_ratio}) is kept at $t = 700$~s, and shifted to later times for higher values of $B_z$ to place it at the beginning of the rapid change phase (continuously increasing derivative), as can be seen in Figure~\ref{fig:aspect_ratio_curve_fit}. $t = \phi_2$~s is chosen so that for $t>\phi_2$~s the derivative drops with time, i.e., the expansion of the CS slows down. $t = \phi_3$~s is kept at 1800~s for all $B_z$, except for the cases where $B_z$ equals 9 and 12~G, since the sheet expansion does not slow down during the full simulation run. Due to this, $\alpha_2$ can only be calculated until $B_z = 6$~G. There is also a significant difference between the values of $\phi_2$ and $\theta_2$ for aspect ratio and $J_{max}$ (see caption of Figures~\ref{fig:etajmax} and \ref{fig:aspect_ratio_curve_fit} for the values of the relevant timestamps).

In Figure~\ref{fig:appendix_vy_max}, $t=\psi_1$~s is kept at $t = 700$~s, and shifted to later times for $B_z$ exceeding 4.5~G. For $B_z>4.5$~G, the beginning of the rapid expansion phase for $|v_y|_{max}$ is preceded by a time interval of minimal variation in $|v_y|_{max}$ from its initial values (see Figure~\ref{fig:max_vy} and \ref{fig:appendix_vy_max}) which upon inclusion deteriorates the quality of fitting. Hence, it is excluded by increasing the value of $\psi_1$ suitably for high $B_z$. $t = \psi_2$~s is chosen to be the time after which the increase in the maximum upflow velocity of the CS slows down.

For $T_{max}$, time-period selection is rather straightforward due to the monotonically increasing nature of the time derivative of $T_{max}$. $t =\chi_1$~s is kept at $t = 700$~s up to $B_z = 4.5$~G. For higher strengths of the guide-field, Figures~\ref{fig:appendix_Tmax} and \ref{fig:max_temp} indicate a time interval where the maximum temperature within the CS stays approximately at 1~MK, i.e., a delay in the beginning of CS heating due to high guide-fields is apparent. For reasons similar to those explaining Figure~\ref{fig:appendix_vy_max}, $\chi_1$ is shifted to later times to account for this effect and improve the overall fitting. $\chi_2$ is kept at 1800~s for all values of guide-field, since there is no slowing down in the increase in the maximum temperature within the CS. For values of $\chi_1$, see the caption of Figure~\ref{fig:max_temp}.

\subsection*{Note on curve fitting methodology}
The temporal derivatives of $J_{\rm max}$, aspect ratio, $|v_y|_{\rm max}$, 
and $T_{\rm max}$ shown in Figures~\ref{fig:appendix_eta_j_max}--\ref{fig:appendix_Tmax} exhibit a near-linear early stage followed by a transition to a higher-order polynomial, visible in Figures~\ref{fig:appendix_eta_j_max} and \ref{fig:appendix_vy_max}.
Fitting the full temporal range with a single polynomial of degree 4 or higher would avoid the ambiguity in the transition timestamp. However, polynomial fits carry no a priori physical motivation for any particular degree, and their additional free parameters reduce the ability to compare trends across guide-field values in a physically interpretable way.
A single power-law fit over the full interval produces systematically poor residuals due to the clear two-phase structure, as noted in Section~\ref{sec:3.2}. This was the initial motivation for the two-window approach. 
The two-window approach approximates the continuous smooth evolution as two piecewise power-law segments, each characterized by a single physically interpretable exponent. The transition timestamp ($\theta_2$, $\phi_2$, etc.) is identified from the derivative plots as the inflection point between the monotonically increasing and decreasing phases, providing a physically motivated rather than arbitrary split point.
A single function of the form $Q = Q_0 + a\,(t - t_0)^{\alpha_1}\,[1 + b\,(t-t_0)^{\alpha_2-\alpha_1}]$ could in principle model the smooth transition, but would introduce four free parameters with limited physical motivation and is difficult to optimize robustly for small datasets spanning two orders of magnitude in time. Appropriately scaled exponential functions could also be utilized.
On a similar note, the second time window in Figure~\ref{fig:appendix_eta_j_max} can be divided into a monotonously decreasing stage and the remaining non-linear stage, which is physically more natural. By fitting the monotonous decreasing stage with exponent $\delta_2$, a representative case of $B_z~=~3$G, we find that $\delta_2$ changes from 1.0027 to 1.0519, and the error in the fitting exponent shifts from 0.0045 to 0.0181. The increasing trend with $B_z$ in $\delta_2$, as observed in Section~\ref{sec:3.2} is preserved, with their values changing only slightly. The errors in $\delta_2$ are, however, between twice and ten times their previously calculated values. Hence, the two-window data-fitting method utilized in this work is a good compromise between choosing physically meaningful transitions and accurately computing power-law exponents.

\section{Evolution of $\langle B_z\rangle$ and plasma inflow speed into the CS} \label{sec:app2}

The figures presented in this appendix serve a complementary role to the  criteria established in Appendix~\ref{sec:app1} for the selection of data windows used in our power-law fitting. Specifically, we demonstrate that the time intervals $\phi_1$--$\phi_2$, $\phi_2$--$\phi_3$ (for $\alpha_1$ 
and $\alpha_2$), and $\theta_1$--$\theta_2$, $\theta_2$--$\theta_3$ (for $\delta_1$ and $\delta_2$) are not chosen arbitrarily, but correspond to phases that can be physically distinguished by the evolution of quantities that are independent diagnostics of the CS dynamics, namely the average guide-field strength within the CS and the inflow velocity Alfvén Mach number into the CS region.


\begin{figure*}[t!]
\centering
\includegraphics[width=\linewidth]{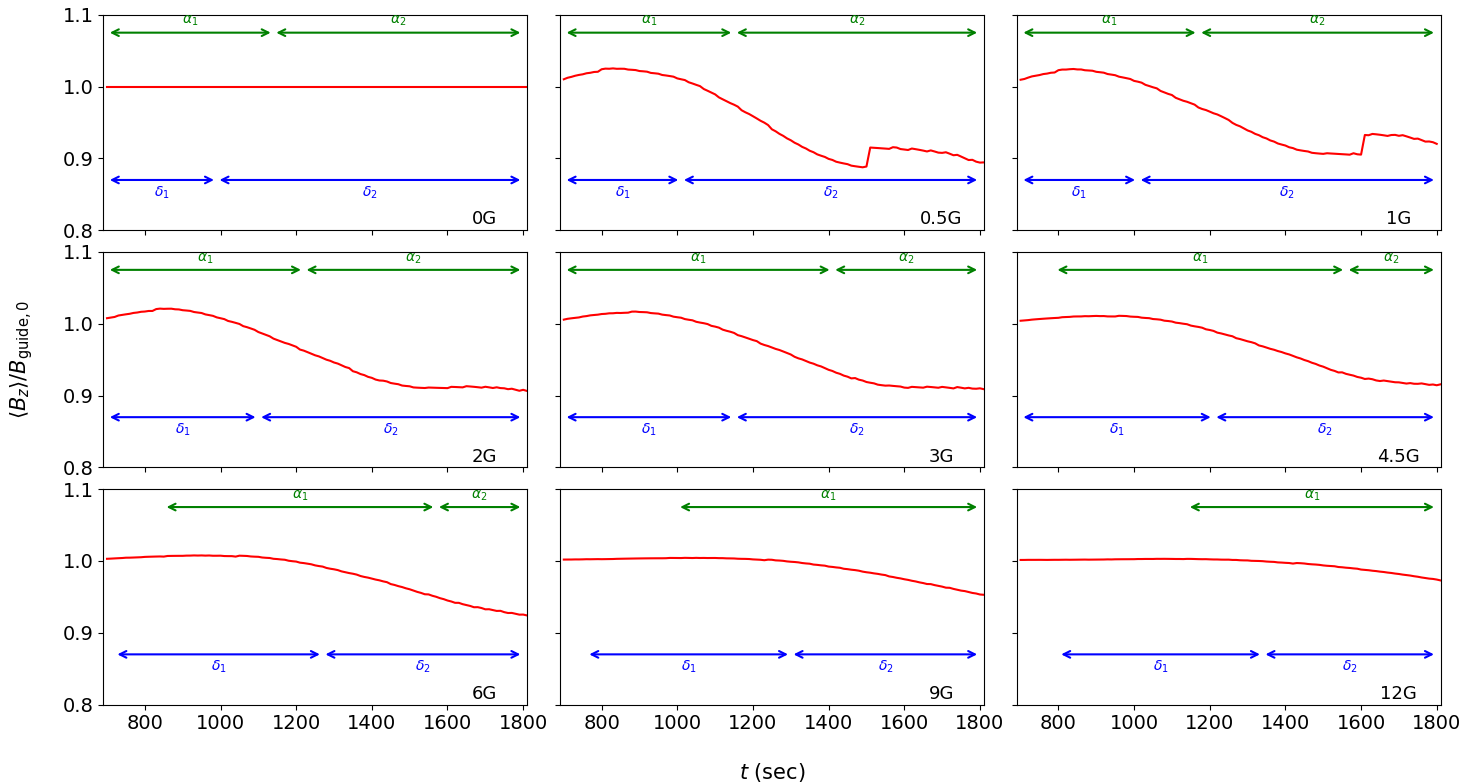}
\caption{Temporal variation of the average magnetic-field component perpendicular to the plane of reconnection relative to the initially set value of guide-field. The average of $B_z$ is taken over the area of the CS as described previously. Green arrowheads denote the data windows over which the power-law exponents for the aspect ratio have an incremental and decremental nature with increasing guide-field, i.e., $\alpha_1$ and $\alpha_2$ respectively. Blue arrowheads denote the data windows over which the power-law exponents for $J_{max}$ have an incremental and decremental nature with increasing guide-field, i.e., $\delta_1$ and $\delta_2$ respectively. 
\label{fig:appendix_bz}
}
\end{figure*}

\subsection*{Evolution of $\langle B_z \rangle / B_{z,0}$}
Figure~\ref{fig:appendix_bz} shows the temporal evolution of the average guide-field component $\langle B_z \rangle$ within the CS, normalized to the initially applied uniform guide-field value $B_{z,0}$. For the zero-guide-field case ($B_{z,0} = 0$~G), normalization is not possible; instead, the absolute value is offset by unity for visual comparison, and the resulting curve remains flat throughout, confirming  that no guide-field component develops within the CS in the absence of an initial guide field.
For all nonzero guide-field cases, the normalized ratio  $\langle B_z \rangle / B_{z,0}$ exhibits a characteristic two-phase evolution. In the initial phase, there is a temporary amplification of $B_z$ within the CS relative to its initial value. This amplification is most pronounced for weaker guide-fields --- reaching up to approximately 4--5\% above unity for $B_{z,0} = 0.5$~G --- and progressively becomes less pronounced with increasing $B_{z,0}$, such that for $B_{z,0} = 12$~G no discernible peak is present and the ratio remains around unity until approximately $t = 1400$~s. 
Following this initial phase, the ratio declines appreciably, reaching approximately 90\% of the initial value for weaker guide fields, with the rate of decline slowing down systematically as $B_{z,0}$ increases. For $B_{z,0} = 12$~G, the decline is minimal, with the ratio dropping only to around 95\% by the end of the simulation. The ratio subsequently flattens for low and intermediate values of $B_{z,0}$, and the value at $t = 1800$~s stays near unity for the higher $B_{z,0}$, consistent with the idea that a stronger guide field is resisting any significant restructuring within the CS.
The correspondence between the data windows and the physical evolution described above is as follows. The $\alpha_1$ window (green arrowheads, Figure~\ref{fig:appendix_bz}) spans the initial amplification peak and the onset of the declining phase. For weaker guide fields, this window starts at $t = 700$~s and captures the full peak; for stronger guide fields, where the ratio remains near unity for an extended period, the window is shifted to later times to coincide with the delayed rapid expansion phase of the CS identified in Appendix~\ref{sec:app1}. The $\alpha_2$ window spans the remainder of the decline phase of the average guide-field strength and the subsequent flattening. The $\delta_1$ and $\delta_2$ windows (blue arrowheads) follow a qualitatively similar correspondence to $\alpha_1$ and $\alpha_2$ but span a somewhat shorter and longer interval respectively relative to their $\alpha$ counterparts. The systematic shift of both the $\alpha_1$ and $\delta_1$ windows with increasing $B_{z,0}$ is physically grounded in the delayed onset of guide-field restructuring within the CS.

\begin{figure*}[ht!]
\centering
\includegraphics[width=\linewidth]{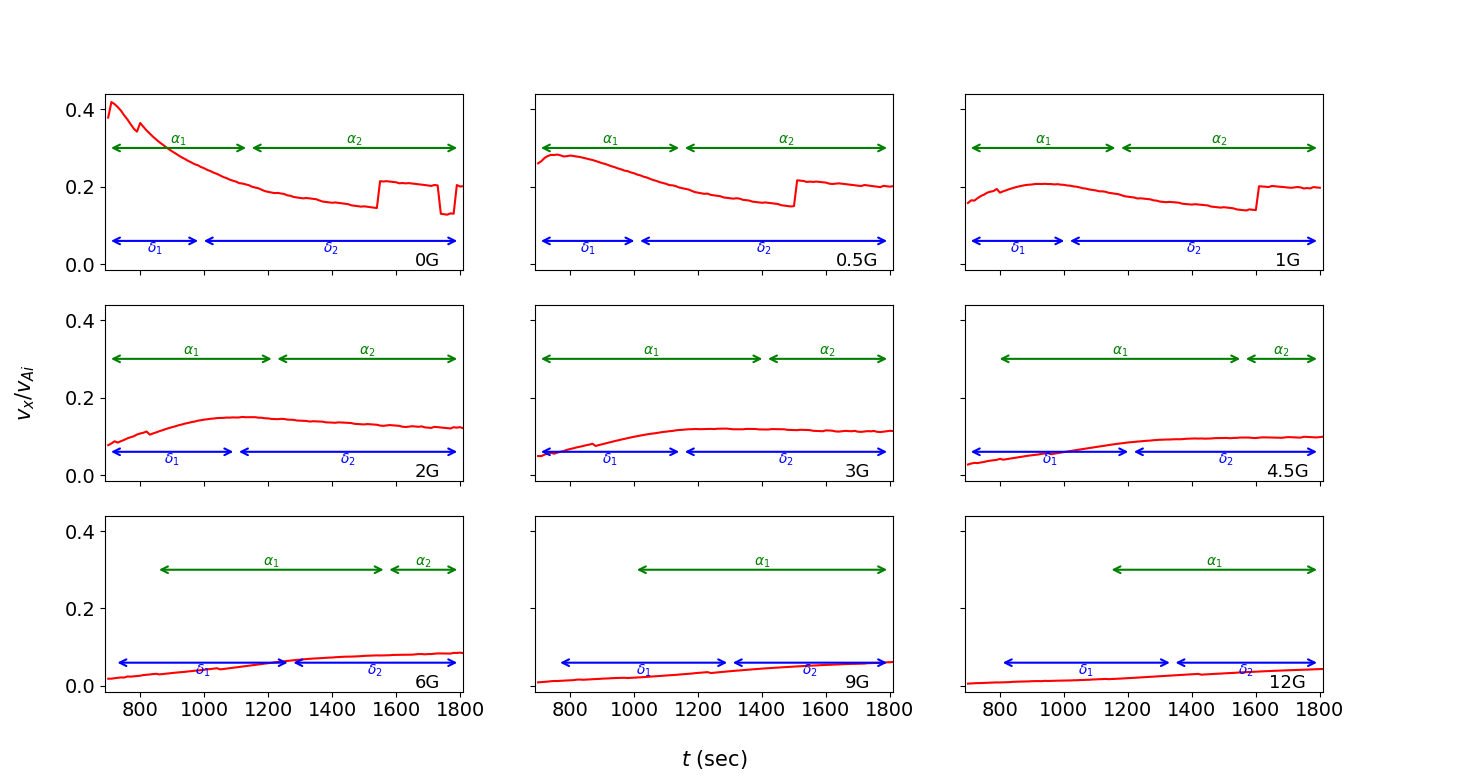}
\caption{Temporal variation of the inflow velocity Alfvén Mach number into the CS. The value of $|v_x|_{inflow, CS}$ is taken by computing the values of $v_x$ at the contours of the CS, using criteria described in Section~\ref{sec:aspectratio}. The inflow velocity is normalized using the local Alfven velocity. Green arrowheads denote the data windows over which the power-law exponents for the aspect ratio have an incremental and decremental nature with increasing guide-field, i.e., $\alpha_1$ and $\alpha_2$ respectively. Blue arrowheads denote the data windows over which the power-law exponents for $J_{max}$ have an incremental and decremental nature with increasing guide-field, i.e., $\delta_1$ and $\delta_2$ respectively. 
\label{fig:appendix_vx}
}
\end{figure*}

\subsection*{Evolution of $|v_x|_{\mathrm{inflow,\,CS}}$}

Figure~\ref{fig:appendix_vx} shows the temporal evolution of the inflow speed of plasma into the CS, $|v_x|_{\mathrm{inflow,\,CS}}$, with respect to the local Alfv\'en speed , $v_{Ai}$, since the magnetic topology keeps changing around the null point as the CS evolves. All relevant quantities are computed at the vertical edges of the considered CS, at the height corresponding to $J_{max}$, as described in Section~\ref{sec:aspectratio}. Direct analysis of inflow speeds displays a monotonously increasing profile across all guide-field values, roughly within the same range (1--8~km~s$^{-1}$). The local Alfv\'en speeds at these locations, however, display a roughly two-stage evolution: first, a negligible change in $v_{Ai}$, followed by a monotonous increase. With increasing $B_z$ values, the stage of near constant $v_{Ai}$ increases in duration, and the following rising stage drops in growth rate. For the $B_z~=~12$~G case, this results in a near constant $v_{Ai}$ for the full run, and for the low guide-field cases (0--1~G), a nearly constantly increasing profile. The value of the local Alfv\'en speed at $t=700$~s increases for increasing $B_z$, as expected. This results in the Alfv\'en Mach number profile of plasma inflow into the CS presented in Figure~\ref{fig:appendix_vx}. The duration of negligible variation in $v_{Ai}$ simply scales down the increasing inflow speed profile, which results in the increasing Mach number profiles prominently observed in Figure~\ref{fig:appendix_vx} for the cases of 0.5--4.5~G. Afterward, the increasing local Alfv\'en speed pushes down the inflow Alfv\'en Mach number. For the higher guide-field cases of 6--12~G, long phases of local Alfv\'en speed constancy are followed by periods of slight growth, which basically allows the Alfv\'en Mach number to grow continuously, albeit in a strongly scaled down fashion, owing to the strong guide fields.

This is physically relatable to the criterion established in Appendix~\ref{sec:app1}, where the span of the aspect-ratio growth exponent $\alpha_1$ is shifted to mark the beginning of the rapid expansion phase, which in Figure~\ref{fig:appendix_vx} overlaps with the increasing Mach number phase (or equivalently, constant $v_{Ai}$ phase), except for 0 and 0.5~G. The $\alpha_2$ window in all cases covers the post-peak (or the increasing $v_{Ai}$ duration) phase. For the highest guide-field cases, its span is reduced due to the rightward displacement of $\phi_1$ reducing the available fitting interval, which corresponds to a near constant local Alfv\'en speed.
For the $\delta_1$ window (blue arrowheads), the correspondence is similar. However, for $B_{z,0} \leq 6$~G, the span of $\delta_1$ encompasses the initial peak and a part of the drop phase, while for higher $B_{z,0}$ it fails to span the peak as the peak is not attained within the simulation runtime anymore. The $\delta_2$ window in all cases (except 9 and 12~G) covers the post-peak Mach number drop and is consistently wider than the corresponding $\alpha_2$ interval. 
$\alpha_1$ and $\delta_1$ roughly reflect the increasing inflows, the growth rate of which decreases with increasing $B_z$, as can be seen prominently in the cases of 1~-~12~G. Similarly, $\alpha_2$ and $\delta_2$ span the decreasing Mach number profile, which for increasing $B_z$ shows an increasing trend in its growth rate. This reflects the conclusions presented in Section~\ref{sec:highlight}.
Taken together, Figures~\ref{fig:appendix_bz} and \ref{fig:appendix_vx} demonstrate that the data windows defined in Appendix~\ref{sec:app1}, using separate numerical criteria for each quantity, correspond to physically meaningful transitions in CS dynamics. 
It should be stated that the sharp transitions observed in Figure~\ref{fig:appendix_bz} corresponding to 0.5 and 1~G and Figure~\ref{fig:appendix_vx} corresponding to 0, 0.5, and 1~G are a consequence of rapid thinning and discretization effects of the domain. The thinnest CSs are obtained in the cases of 0, 0.5, and 1~G, the width of which approaches the cell width in the $x$-dimension. A shift in the contour of one or two cells causes a discrete jump in the computed CS width, which propagates to the averaged quantities as an abrupt bump. This sensitivity is unavoidably intrinsic to any threshold-based CS detection routine applied to a dynamically evolving thin sheet.

\section{Comparison between thermal and guide-field induced in-plane dynamics}\label{app:bzvsp}
\begin{figure*}[]
\centering
\includegraphics[width=\linewidth]{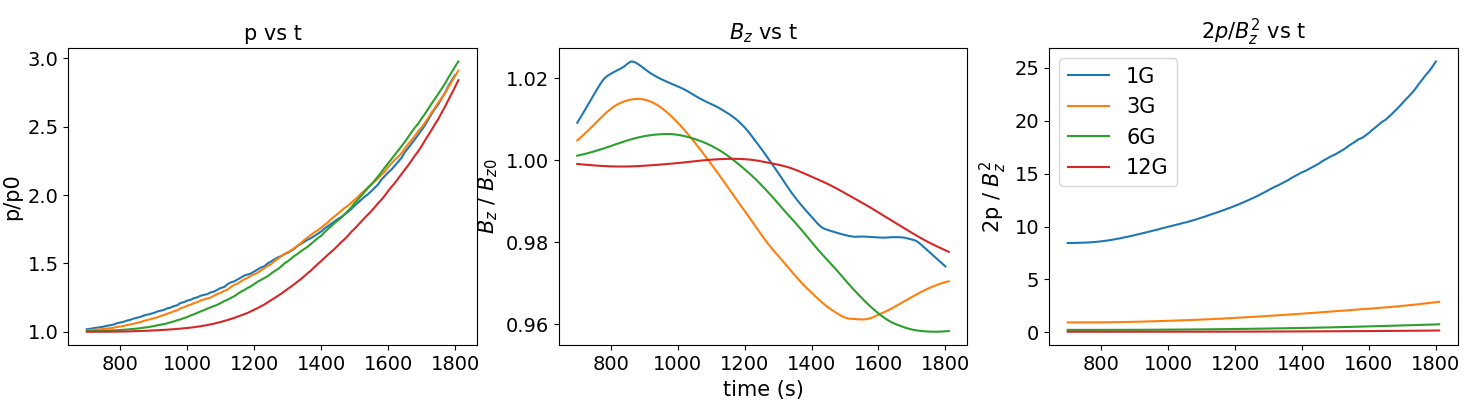}
\caption{Temporal evolution of thermal pressure within the CS (left), guide-field strength within the CS (center) and measurement of contribution between thermal and guide-field pressure in the pressure-gradient term (right).
\label{fig:appendix_prsVsBzSqr}
}
\end{figure*}

In the 2.5D approximation, the variation of the guide-field $B_z$ contributes to the pressure gradient term alongside the thermal pressure to influence in-plane dynamics. In the corona, plasma pressure can generally be ignored, but when considering Ohmic heating this is no longer necessarily true. In this section, we attempt to quantify the relative contribution between thermal and guide-field pressure, and what implications these findings carry for the work presented in this paper.
To this end, we extract two quantities from our simulations: thermal pressure and guide-field strength at the location of maximum current density within the CS. The ratio of the thermal pressure to the guide-field contribution to the in-plane pressure is also computed.
Figure~\ref{fig:appendix_prsVsBzSqr} shows the variation of these quantities with time for the cases of $B_z~=~1,~3,~9,~12$~G. Not all values of $B_z$ are considered for clarity, but follow trends that can be interpolated from the representative values in Figure~\ref{fig:appendix_prsVsBzSqr}.
The first and second panels (from left to right) in Figure~\ref{fig:appendix_prsVsBzSqr} show the variation of thermal pressure and guide-field strength inside the CS with respect to the initially set values. Thermal pressure varies between the initial and at most three times the initial pressure, with the only difference across guide-fields being the growth rate, which decreases with increasing $B_z$. This is akin to the variation of $T_{max}$ in Figure~\ref{fig:appendix_Tmax}.
The indicative guide-field strength within the CS with respect to the initially set value (different for each case) varies in a similar fashion to Figure~\ref{fig:appendix_bz}. We observe that the guide-field strength varies between $\approx~0.96~-~1.03$ times the initial guide-field strength.
The ratio of the thermal to guide-field pressure varies as such: 34--111 for 0.5~G, 8--25 for 1~G, 2.1--6.3 for 2~G, 0.93--2.9 for 3~G, 0.4--1.3 for 4.5~G, 0.2--0.8 for 6~G, 0.1--0.3 for 9~G, and 0.05--0.17 for 12~G. For the 2D case of $B_z~=~0$~G, all dynamics are naturally dominated by the thermal pressure. For 0.5--2~G, thermal pressure still dominates the dynamics, but to a lesser extent than in the 2D counterparts. For 3--4.5~G, the guide-field plays a more determining role compared to the thermal pressure in the gradient term, which transitions to thermal pressure making a comparable contribution during the later stages of CS evolution. For 6--12~G, the guide-field term dominates the pressure-gradient term, and continues to do so over the course of complete evolution of the CS. The evolution of $2~p~/~B_z^2$ across the simulation runtime is governed by thermal pressure variations, owing to the much weaker variations in guide-field strength. Hence, the relative contribution between the thermal and guide-field pressures is determined primarily by the initial guide-field strength, which stays roughly within the same order throughout the runtime of the simulation. 

An important implication concerns the applicability of these results 
to the real solar corona, where the resistivity is approximately 
eight orders of magnitude smaller than that adopted here 
(Section~\ref{subsec:ibc}). Ohmic heating scales as $\eta J^2$; 
at realistic coronal resistivities, the thermal pressure enhancement 
within the CS would be substantially smaller for the same CS geometry. 
Consequently, the thermal-to-guide-field pressure ratio would be 
shifted significantly toward guide-field dominance across all $B_z$ 
values considered here. In the real corona, even the $B_z = 0.5$--$2$~G 
cases --- which are thermally dominated in our simulations --- would 
likely operate in a guide-field-dominated or at least transitional 
regime. A systematic study of the thermal-to-guide-field pressure 
ratio as a function of Lundquist number, achieved through a 
resistivity parameter-study at fixed guide-field strength, would be required to 
quantify this effect rigorously and could be an 
extension of the present work.

\FloatBarrier 
\clearpage

\end{appendix}

\end{document}